\documentclass[11pt,a4paper]{article}
\pdfoutput=1

\usepackage{jheppub}
\usepackage{amsmath}
\usepackage{amssymb}
\usepackage{graphics}
\usepackage[active]{srcltx}
\usepackage{pdfsync}
\usepackage{shuffle}
\usepackage{slashed}
\usepackage{hyperref}
\usepackage{subfigure}

\newcommand{\bea}{\begin{eqnarray}}
\newcommand{\eea}{\end{eqnarray}}
\newcommand{\nn}{\nonumber \\}

\def\W #1{\widetilde{#1}}

\def\eref#1{(\ref{#1})}

\def\a{{\alpha}}

\def\b{{\beta}}

\allowdisplaybreaks

\title{Soft theorems of tree-level ${\rm Tr}(\phi^3)$, YM and NLSM amplitudes from $2$-splits}
\author[a]{Kang Zhou}

\affiliation[a]{Center for Gravitation and Cosmology, College of Physical Science and Technology, Yangzhou University,\\
No.180, Siwangting Road, Yangzhou, 225009, P.R. China}

\date{\today}
\abstract{In this paper, we extend the method proposed in \cite{Arkani-Hamed:2024fyd} for deriving soft theorems of amplitudes, which relies exclusively on factorization properties including conventional factorizations on physical poles, as well as newly discovered $2$-splits on special loci in kinematic space. Using the extended approach, we fully reproduce the leading and sub-leading single-soft theorems for tree-level ${\rm Tr}(\phi^3)$ and Yang-Mills (YM) amplitudes, along with the leading and sub-leading double-soft theorems for tree-level amplitudes of non-linear sigma model (NLSM). Furthermore, we establish universal representations of higher-order single-soft theorems for tree-level ${\rm Tr}(\phi^3)$ and YM amplitudes in reduced lower-dimensional kinematic spaces. All obtained soft factors maintain consistency with momentum conservation; that is, while each explicit expression of the resulting soft behavior may changes under re-parameterization via momentum conservation, the physical content remains equivalent. Additionally, we find two interesting by-products: First, the single-soft theorems of YM amplitudes and the double-soft theorems of NLSM, at leading and sub-leading orders, are related by a simple kinematic replacement. This replacement also transmutes gauge invariance to Adler zero.
Second, we obtain universal sub-leading soft theorems for the resulting pure YM and NLSM currents in the corresponding $2$-splits.}

\begin{document}

\maketitle \flushbottom

\section{Introduction}
\label{sec-intro}

In recent years, three novel types of factorizations of tree-level amplitudes, termed smooth splitting, factorization near hidden zero, and $2$-split, have been discovered successively \cite{Cachazo:2021wsz,Arkani-Hamed:2023swr,Cao:2024gln,Cao:2024qpp,Arkani-Hamed:2024fyd}.
Unlike the conventional tree-level factorization, which states the behavior of residue on any physical pole, these new factorizations occur on special loci in kinematic space without requiring residue evaluation. It has been shown that among the three types, the $2$-split is the most fundamental one, as both smooth splitting and factorization near zero arise from further specializations of consecutive $2$-splits \cite{Cao:2024qpp}. As demonstrated in \cite{Cao:2024gln,Cao:2024qpp}, the $2$-split is a universal phenomenon, with a wide range of particle and string amplitudes exhibiting this behavior.
Whereas the conventional factorization produces two on-shell lower-point amplitudes, the $2$-split yields two amputated currents, each with an off-shell leg. These novel factorizations and associated hidden zeros have attracted extensive attention, see in  \cite{Zhang:2024iun,Zhang:2024efe,Rodina:2024yfc,Bartsch:2024amu,Li:2024qfp,Zhou:2024ddy,Huang:2025blb,Feng:2025ofq}.

As is well known, the usual factorization on poles enables effective computational techniques, as well as novel formalisms and insights, such as the famous Britto-Cachazo-Feng-Witten (BCFW) on-shell recursion relation and the amplituhedron \cite{Britto:2004ap,Britto:2005fq,Arkani-Hamed:2013jha,Arkani-Hamed:2017mur}. It is natural to expect that the new class of factorizations will also lead to significant physical implications and applications.
For instance, hidden zeros and their associated factorizations provided the crucial clue for discovering the simple "$\delta$-shift" of kinematic variables, which preserves zeros and factorizations. This marvelous shift unifies the ${\rm Tr}(\phi^3)$, non-supersymmetric Yang-Mills (YM), and non-linear sigma model (NLSM) amplitudes into a single stringy curve integral \cite{Arkani-Hamed:2023jry,Arkani-Hamed:2024nhp}, in the story of "surfaceology" which represents a major advance in recent studies of scattering amplitudes \cite{Arkani-Hamed:2023lbd,Arkani-Hamed:2023mvg,Arkani-Hamed:2023jry,Arkani-Hamed:2024nhp,Arkani-Hamed:2024tzl,Arkani-Hamed:2024vna, Arkani-Hamed:2024yvu,Arkani-Hamed:2024nzc,Arkani-Hamed:2024pzc}. Furthermore, as demonstrated in \cite{Arkani-Hamed:2024fyd}, the novel $2$-split also provides a new approach for deriving universal soft theorems of amplitudes.

This paper focuses on deriving soft theorems for tree-level amplitudes by utilizing $2$-splits. Soft theorems describe a universal factorization behavior when one or more external massless particles become soft, that is, their momenta are scaled as
$k_i=\tau\hat{k}_i$, with $\tau\to 0$. In such soft limit, a wide range of amplitudes at specific orders of $\tau$ factorize into a universal soft factor and a lower-point amplitude excluding the soft particles. For instance, when an external gluon $s$ is taken soft, the color-ordered tree-level YM amplitude factorizes as
\bea
{\cal A}_{\rm YM}(1,\cdots,n,s)=\Big({1\over\tau}\,S^{(0)}_g+S^{(1)}_g\Big)\,{\cal A}_{\rm YM}(1,\cdots,n)+\cal O(\tau)\,,~~\label{soft-example}
\eea
where $S^{(0)}_g$ and $S^{(1)}_g$ are universal leading and sub-leading soft factors, respectively. Here, "universal" means their forms are independent of the number of external particles. As demonstrated in a series of works, the soft behaviors can be exploited to construct tree amplitudes without relying on traditional Lagrangians or Feynman rules \cite{Nguyen:2009jk,Boucher-Veronneau:2011rwd,Rodina:2018pcb,Ma:2022qja,Cheung:2014dqa,Cheung:2015ota,Luo:2015tat,Elvang:2018dco}. For example, it allows us to access the universal expansions of tree amplitudes for a large variety of theories, via an on-shell recursive scheme that invert soft limits \cite{Zhou:2022orv,Wei:2023yfy,Hu:2023lso,Du:2024dwm,Zhou:2024qjh,Zhou:2024qwm}. These expansions hold for arbitrary numbers of external states, simplifying amplitude computations to evaluating coefficients (with prescribed rules) and basis.

However, such constructions discussed above presuppose known soft theorems. In general, deriving soft theorems for multiple distinct theories, via a universal methodology without prior knowledge of amplitudes, is highly non-trivial. The method in \cite{Arkani-Hamed:2024fyd}, leveraging $2$-splits and conventional factorizations, offers a new approach to tackle this challenge. Unfortunately, while it fully determines leading-order soft theorems via $2$-splits alone, higher-order results derived via the method in \cite{Arkani-Hamed:2024fyd} are restricted to lower-dimensional kinematic subspaces, i.e., they are incomplete.
Nevertheless, most inverse-soft methods in the literature require complete sub-leading soft factors (e.g., $S^{(1)}_g$
in \eref{soft-example}) to fully construct amplitudes. Yet, the method in \cite{Arkani-Hamed:2024fyd} cannot supply the complete set of needed soft factors.

To address this gap, we extend the method of \cite{Arkani-Hamed:2024fyd} to derive complete sub-leading soft theorems. Unlike \cite{Arkani-Hamed:2024fyd} which uses one $2$-split, our method employs two. Furthermore, while the treatment for YM and NLSM in \cite{Arkani-Hamed:2024fyd} depends on the "$\delta$-shift" formalism of surfaceology, our method is independent of this framework. By combining
(i) conventional factorizations on poles, and
(ii) two distinct $2$-splits on different kinematic loci,
we obtain full sub-leading single-soft factors for ${\rm Tr}(\phi^3)$/YM amplitudes and double-soft factors for NLSM at tree-level. Here, "single-soft" denotes one soft particle, while "double-soft" involves two.
Although we focus on these three theories, our method is generalizable to any amplitude exhibiting 2-splits, as it avoids theory-specific properties (e.g., gauge invariance or Adler zero). To systematize our approach, we also discuss the derivation of leading soft theorems, via two distinct methods: one based on conventional factorizations and the other utilizing novel $2$-splits. Remarkably, while the conventional factorization approach is limited to ${\rm Tr}(\phi^3)$ and YM amplitudes, the $2$-split method demonstrates broader applicability, successfully extending to NLSM amplitudes as well.

An interesting observation is, the single-soft theorems for Yang-Mills (YM) amplitudes (at leading/sub-leading orders) and the double-soft theorems for NLSM amplitudes (at the same orders) are interrelated through a simple kinematic replacement: $\epsilon_s\to k_{s_1}-k_{s_2}$, $k_s\to k_{s_1}+k_{s_2}$, where $\epsilon_s$ and $k_s$ denote the polarization vector and momentum of the soft gluon, while $k_{s_1}$ and $k_{s_2}$ are momenta of two soft pions. This replacement maps gauge invariance in YM to Adler zero in NLSM. The construction which builds gluon polarizations from scalar momenta, plays a pivotal role in the "surfaceological" description of gluon amplitudes, known as scalar-scaffolded gluons \cite{Arkani-Hamed:2023jry}.

While the sub-leading soft theorems derived through our approach are not automatically guaranteed to hold in the full kinematic space (being initially valid only in the subspace defined by two specific $2$-splits), we propose a self-contained verification scheme that confirms their validity across the complete kinematic space, without requiring comparison with known forms of soft factors in existing literature. For ${\rm Tr}(\phi^3)$ amplitudes, the validity of sub-leading soft factor is confirmed by analyzing solely the power of soft scale parameter $\tau$. For more complicated YM and NLSM ones, we cross-check results via additional $2$-splits at distinct kinematic loci. This process also yields universal sub-leading soft theorems for the pure currents of YM and NLSM in their respective $2$-splits. More explicitly, each pure YM or pure NLSM current carries an off-shell external leg $\kappa'$ with combinatorial momentum $k_{\kappa'}=k_k+k_A$, where $k_A$ is a sum of some on-shell massless momenta. When all components of $k_A$ become soft, the current factorizes universally as
\bea
{\rm Current}\,\rightarrow\,{\rm Soft~factor}\,\times\,{\rm Amplitude}\,.
\eea

Our method for deriving sub-leading soft theorems naturally extends to higher orders. For both ${\rm Tr}(\phi^3)$ and Yang-Mills (YM) theories, we obtain universal representations of soft theorems at arbitrary $m^{\rm th}$ order (where $m$ is any positive integer) through this approach. However, similar to the results in \cite{Arkani-Hamed:2024fyd}, these general soft theorems are restricted to reduced lower-dimensional kinematic subspaces and therefore do not represent complete formulations.
This partial result parallels previous work on extending soft theorems to arbitrary orders through appropriate projections, as investigated in \cite{Hamada:2018vrw,Li:2018gnc}. To illustrate, consider the case of a single soft photon $s$, which exhibits the following $m^{\rm th}$ order soft behavior:
\bea
\Omega_{\mu\a_1\cdots\a_m}\,\partial_{k_s}^{\a_1}\cdots\partial_{k_s}^{\a_m}\,{\cal A}_{n+1}^\mu=\sum_{j=-1}^\infty\,S^\mu_{j,\nu}\,{\cal A}_{n}^\nu\,,~~\label{soft-extend}
\eea
where ${\cal A}_n^\mu$ is the amplitude with one polarization vector stripped off, and $\Omega_{\mu\a_1\cdots\a_l}$ is a totally symmetric tensor.
The higher order soft theorems presented in this paper share a similar projection property with those in \cite{Hamada:2018vrw,Li:2018gnc}, where in our case the projection is performed onto specific subspaces determined by pairs of 2-split loci.

Since amplitude expressions can always be re-parameterized using momentum conservation, we require that any soft factor derived via our method, when viewed as an operator acting on lower-point amplitudes, must satisfy the essential requirement that any momentum-conserving re-parameterization of lower-point amplitude yields equivalent soft behavior. We call such property the consistency with momentum conservation. We will verify that  our general soft theorems in subspaces satisfy such consistency condition, by defining the momentum conservation operator, and analyzing its commutator with soft operator.

The reminder of this paper is organized as follows. In section \ref{sec-2split}, we briefly introduce the $2$-split and its key properties for latter use. Then, in section \ref{sec-phi3}, we derive single-soft theorems (leading/sub-leading/higher-order) for tree-level ${\rm Tr}(\phi^3)$ amplitudes, by exploiting usual factorizations and $2$-splits. In section \ref{sec-YM}, we generalize the method in section \ref{sec-phi3} to the YM case, and establish the analogous single-soft theorems. In section \ref{sec-NLSM}, we adapt the approach to derive leading/sub-leading double-soft theorems for NLSM. Finally, in section \ref{sec-summary}, we end with a summary and discussion.

\section{$2$-splits of ${\rm Tr}(\phi^3)$, YM and NLSM amplitudes}
\label{sec-2split}

For readers' convenience, in this section we rapidly review the novel factorization behavior called $2$-split, for tree-level ${\rm Tr}(\phi^3)$, YM and NLSM amplitudes
\cite{Cao:2024gln,Cao:2024qpp,Arkani-Hamed:2024fyd}.

\begin{figure}
  \centering
   \includegraphics[width=13cm]{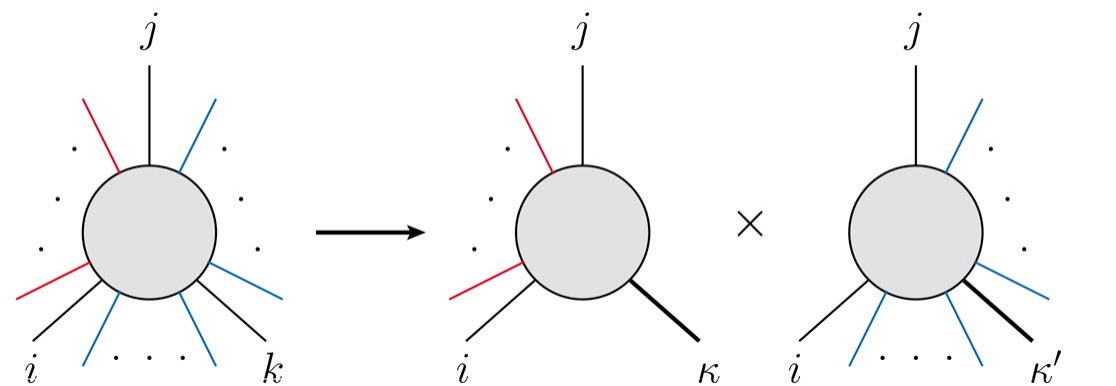}\\
  \caption{$2$-split of tree-level ${\rm Tr}(\phi^3)$/YM/NLSM amplitudes, where legs in the set $A$ are represented in red, while those in the set $B$ are represented in blue. The bold lines represent off-shell particles.}\label{split-general}
\end{figure}

The ${\rm Tr}(\phi^3)$ theory describes the cubic interaction of colored massless scalars, with the following Lagrangian
\bea
{\cal L}_{{\rm Tr}(\phi^3)}={\rm Tr}(\partial\phi)^2+g\,{\rm Tr}(\phi^3)\,,
\eea
where $\phi$ is an $N\times N$ matrix, with two indices in the fundamental and anti-fundamental representations of
$SU(N)$. Each color ordered tree-level amplitude of this theory, with coupling constants stripped off, consists solely of propagators for massless scalars, and is planer with respect to the ordering. The $2$-split of such amplitude can be achieved as follows. For the $n$-point amplitude ${\cal A}_{\rm Tr(\phi^3)}(1,\cdots,n)$, one can choose three external legs $i$, $j$, $k$, with $i<j<k$, and divide the remaining legs into two sets $A=\{i+1,\cdots,j-1\}$ and $B=\{j+1,\cdots,i-1\}\setminus k$, then the amplitude factorizes into two amputated currents with an off-shell leg for each of them,
\bea
{\cal A}_{\rm Tr(\phi^3)}(1,\cdots,n)\,\xrightarrow[a\in A\,,\,b\in B]{k_a\cdot k_b=0}\,{\cal J}_{{\rm Tr}(\phi^3)}(i,\cdots,j,\kappa)\,\times\,{\cal J}_{{\rm Tr}(\phi^3)}(j,\cdots,\kappa',\cdots,i)\,,~~\label{2split-phi3-general}
\eea
on the locus $k_a\cdot k_b=0$ in kinematic space, where $a\in A$, $b\in B$. Such behavior is illustrated in Fig.\ref{split-general}.
In the above, the on-shell external leg $k$ is turned to off-shell ones $\kappa$
and $\kappa'$ in two resulting currents. The momentum conservation forces $k_\kappa=k_k+k_B$ and $k_{\kappa'}=k_k+k_A$ respectively, where $k_A=\sum_{a\in A}k_a$ encodes the
total momentum carried by legs in $A$, while $k_B=\sum_{b\in B}k_b$ denotes the
total momentum carried by legs in $B$.

The tree-level YM and NLSM amplitudes have the similar $2$-splits. The $2$-split of YM is given as
\bea
{\cal A}_{\rm YM}(1,\cdots,n)\,\xrightarrow[a\in A\,,\,b\in B]{\eref{kine-condi-YM}}\,{\cal J}_{{\rm Tr}(\phi^3)\oplus {\rm YM}}(i_\phi,A_g,j_\phi,\kappa_\phi)\,\times\,\epsilon_k\cdot{\cal J}_{\rm YM}(j,\cdots,\kappa',\cdots,i)\,,~~\label{2split-YM-general}
\eea
with the following constraints on external momenta and polarization vectors,
\bea
\{\epsilon_a\,,\,k_a\}\,\cdot\,\{\epsilon_b\,,\,k_b\,,\,\epsilon_c\}=0\,,~~~~{\rm for}~a\in A\,,~b\in B\,,~c\in\{i,j,k\}\,,~~\label{kine-condi-YM}
\eea
where $\pmb\a\cdot \pmb\b=0$ for two sets $\pmb\a$ and $\pmb\b$ means $\a\cdot\b=0$ for any $\a\in\pmb\a$ and $\b\in\pmb\b$.
In \eref{2split-YM-general}, the left resulting current is the mixed one of ${\rm Tr}(\phi^3)\oplus$YM theory. The external legs $i$,
$j$ and $\kappa$ of this left current are scalars of ${\rm Tr}(\phi^3)$ theory, as labeled by the subscript $\phi$, while legs in the set $A$ are gluons labeled by $g$. The right pure YM current is a vector one, and is contracted with the polarization $\epsilon_k$. If we replace the condition \eref{kine-condi-YM} by $\{\epsilon_a,k_a,\epsilon_c\}\,\cdot\,\{\epsilon_b,k_b\}=0$, then the left current is turned to the pure YM one, while the right current should be understood as the mixed one which contains external ${\rm Tr}(\phi^3)$ scalars $i$, $j$, $\kappa'$, and external gluons in $B$. The $2$-split of NLSM reads
\bea
{\cal A}_{\rm NLSM}(1,\cdots,n)\,\xrightarrow[a\in A\,,\,b\in B]{k_a\cdot k_b=0}\,{\cal J}_{{\rm Tr}(\phi^3)\oplus{\rm NLSM}}(i_\phi,A_p,j_\phi,\kappa_\phi)\,\times\,{\cal J}_{\rm NLSM}(j,\cdots,\kappa',\cdots,i)\,.~~\label{2split-NLSM-general}
\eea
In the above, the left resulting current is the mixed one of ${\rm Tr}(\phi^3)\oplus$NLSM. The external legs $i$, $j$, $\kappa$ of this current are scalars of ${\rm Tr}(\phi^3)$ theory, while those in $A$ are pions. Since all of interactions in the NLSM theory are for even multiplicities, the $2$-split \eref{2split-NLSM-general} holds as long as that the set $A$ contains the even number of elements. Suppose the set $A$ contains the odd number of particles, the left current becomes the pure NLSM one, while the right current should be interpreted as the mixed one which contains external ${\rm Tr}(\phi^3)$ scalars $i$, $j$, $\kappa'$, as well as pions in $B$.

\begin{figure}
  \centering
   \includegraphics[width=13cm]{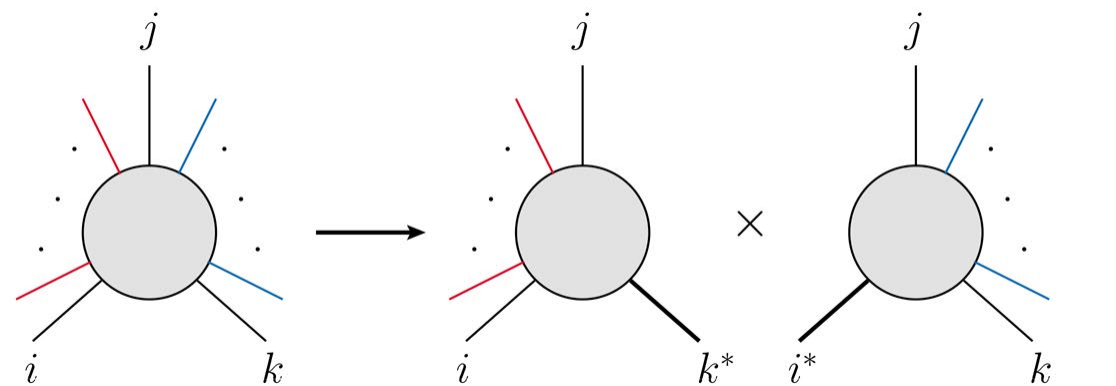}\\
  \caption{$2$-split for tree amplitudes with nearby $k$ and $i$ in the color ordering, with another choice of off-shell legs which are represented by bold lines. Legs in the set $A$ are represented in red, while those in the set $B$ are represented in blue.}\label{split-special}
\end{figure}

In \eref{2split-phi3-general}, \eref{2split-YM-general} and \eref{2split-NLSM-general}, the on-shell leg $k$ is replaced by off-shell legs $\kappa$ and $\kappa'$ in two currents, this is the convention in \cite{Cao:2024qpp}: legs $i$ and $j$ are chosen to be on-shell in both currents thus the remaining $\kappa/\kappa'$ must be off-shell, due to momentum conservation. However, as pointed out in \cite{Cao:2024qpp}, one can also equally make other choices of off-shell legs. In this paper, the special choice of $i$, $j$ and $k$
that $k$ and $i$ (or $j$ and $k$) are adjacent in the ordering, i.e., $i=k+1$ (or $k=j+1$), plays the crucial role. We focus on the case with nearby $k$ and $i$, but all discussions in the rest of this section hold for the case with adjacent $j$ and $k$. Based on our technic for constructing soft factors, for the situation with consecutive $k$ and $i$ it is more convenient to chose off-shell legs as in Fig. \ref{split-special}, namely, the off-shell leg in the left current is $k^*$, while that in the right current is $i^*$. According to the above choice, $2$-splits of ${\rm Tr}(\phi^3)$, YM and NLSM amplitudes are expressed as
\bea
{\cal A}_{\rm Tr(\phi^3)}(1,\cdots,n)\,\xrightarrow[a\in A\,,\,b\in B]{k_a\cdot k_b=0}\,{\cal J}_{{\rm Tr}(\phi^3)}(i,\cdots,j,k^*)\,\times\,{\cal J}_{{\rm Tr}(\phi^3)}(j,\cdots,k,i^*)\,,~~\label{2split-phi3-special}
\eea
\bea
{\cal A}_{\rm YM}(1,\cdots,n)\,\xrightarrow[a\in A\,,\,b\in B]{\eref{kine-condi-YM}}\,{\cal J}_{{\rm Tr}(\phi^3)\oplus {\rm YM}}(i_\phi,A_g,j_\phi,k^*_\phi)\,\times\,\epsilon_i\cdot{\cal J}_{\rm YM}(j,\cdots,k,i^*)\,,~~\label{2split-YM-special}
\eea
and
\bea
{\cal A}_{\rm NLSM}(1,\cdots,n)\,\xrightarrow[a\in A\,,\,b\in B]{k_a\cdot k_b=0}\,{\cal J}_{{\rm Tr}(\phi^3)\oplus{\rm NLSM}}(i_\phi,A_p,j_\phi,k^*_\phi)\,\times\,{\cal J}_{\rm NLSM}(j,\cdots,k,i^*)\,.~~\label{2split-NLSM-special}
\eea
The off-shell momenta can be determined as $k^*_k=k_k+k_B$ and $k^*_i=k_k+k_A$, due to momentum conservation.

In $2$-splits \eref{2split-phi3-special}, \eref{2split-YM-special} and \eref{2split-NLSM-special}, suppose we use momentum conservation to replace all $k^*_i$ by $-(k_j+k_k+k_B)$, then the off-shell right currents share the same expressions with the associated on-shell amplitudes. This property can be observed via the CHY definition of such currents in \cite{Cao:2024qpp}, and can also be seen from the analysis based on Feynman diagrams \cite{Zhou:2024ddy}. Using this property, one can immediately see the advantage of the configuration in Fig. \ref{split-special} for studying soft behaviors. Consider the multi-soft limit that particles in $A$ are taken to be soft, namely,
\bea
k_a\to\tau\hat{k}_a\,,~~~~{\rm with}~\tau\to 0\,.~~\label{multi-soft}
\eea
If $k$ and $i$ are nearby, and the off-shell leg in the right current is chosen to be $i^*$, one can straightforwardly get the behavior of such current at each order, by expanding with $\tau$. Taking the ${\rm Tr}(\phi^3)$ amplitudes as the example. As mentioned before, if we use momentum conservation to remove all $k^*_i$ in ${\cal J}_{{\rm Tr}(\phi^3)}(j,\cdots,k,i^*)$, then this current is expressed the same as the on-shell amplitude. Therefore, if we insert $k^*_i=k_i+k_A$ back, and take the multi-soft limit \eref{multi-soft}, the expansion with $\tau$ can be obtained as
\bea
{\cal J}_{{\rm Tr}(\phi^3)}(j,\cdots,k,i^*)={\cal A}_{{\rm Tr}(\phi^3)}(j,\cdots,k,i)+\sum_{m=1}^\infty\,{\tau^m\over m!}\,(\hat{k}_A\cdot\partial_{k_i})^m\,{\cal A}_{{\rm Tr}(\phi^3)}(j,\cdots,k,i)\,,~~\label{expan-multi}
\eea
where $\hat{k}_A=k_A/\tau$. In the above expansion, the leading term is obvious. Terms at higher orders are derived via the observation that $k_A$ enters the current only through the combinatorial momentum $k^*_i=k_i+k_A$. The expansion \eref{expan-multi} means the the multi-soft behavior of the current ${\cal J}_{{\rm Tr}(\phi^3)}(j,\cdots,k,i^*)$ at any order can be easily captured. The same argument holds for YM and NLSM amplitudes, since they also share the same expressions with on-shell amplitudes if removing $k^*_i$ via momentum conservation.

However, the configuration with un-nearby $k$ and $i$ does not have such luxury, the right current depends on the soft scale parameter $\tau$ via the more complicated manner. Let us give an example for such complexity. For each Feynman diagram, one can always find three lines starting from $i$, $j$ and $k$, and meeting at the vertex $v$. We denote these lines as $L_{i,v}$, $L_{j,v}$ and $L_{k,v}$. As can be seen in examples in \cite{Cao:2024qpp} and discussions in \cite{Zhou:2024ddy}, for un-nearby $k$ and $i$, propagators along the line $L_{\kappa'v}$, which contribute to the right current, should acquire an effective mass. Such effective mass depends on the total momentum $k_A$ from the set $A$, therefore depends on $\tau$. This feature causes additional complexity when expanding the current with $\tau$. However, if $k$ and $i$ are adjacent, the effective mass will not occur.

Due to the above reason, in subsequent sections we will use the configuration in Fig. \ref{split-special}, as well as the associated $2$-splits in \eref{2split-phi3-special}, \eref{2split-YM-special} and \eref{2split-NLSM-special}, to establish soft theorems and soft factors. Meanwhile, we will use the more general $2$-splits \eref{2split-phi3-general}, \eref{2split-YM-general} and \eref{2split-NLSM-general} with un-nearby $k$ and $i$, to verify our results at the sub-leading order.

\section{Single-soft theorems of ${\rm Tr}(\phi^3)$ amplitudes}
\label{sec-phi3}

In this section, we study the single-soft theorems of tree-level ${\rm Tr}(\phi^3)$ amplitudes, with only one external scalar goes soft in each amplitude.
We encode the soft particle as $s$, and parameterize its momentum as $k_s\to\tau\hat{k}_s$, with $\tau\to 0$. Correspondingly, we also parameterize Mandelstam variables that are linear in $\tau$ as $s_{ns}\to\tau\hat{s}_{ns}$, $s_{s1}\to\tau\hat{s}_{s1}$, since $s_{ns}=2k_s\cdot k_n$, $s_{s1}=2k_s\cdot k_1$.

Our derivation employs two basic ingredients, the first one is Feynman diagrams containing divergent propagators in the soft limit, while the second one is the $2$-split in \eref{2split-phi3-special} where $A$ is chosen to be $A=\{s\}$, along with the analogous $2$-split with adjacent $j$ and $k$. The soft theorem at the leading order can be derived using either ingredient alone, as demonstrated in subsection \ref{subsec-leading-phi3}. At this order, the soft limit automatically satisfies the split kinematic $k_s\cdot k_b=0$, ensuring that each $2$-split properly captures the well defined leading soft behavior. In subsection \ref{subsec-subleading-phi3}, we derive the sub-leading soft behavior by combining both ingredients, and argue the correctness of the result by analyzing the order of $\tau$. Subsequently, in Subsection \ref{subsec-higher-phi3}, we extend this method to higher orders. However, the resulting universal form of the higher-order soft factor is incomplete, being valid only in the lower-dimensional subspace of the full kinematic space defined by the $2$-split conditions. Finally, in Subsection \ref{subsec-check-phi3}, we verify that the higher-order soft theorem obtained for this subspace remains consistent with momentum conservation.

\subsection{Leading order}
\label{subsec-leading-phi3}

In this subsection, we derive the simplest leading soft theorem of tree-level ${\rm Tr}(\phi^3)$ amplitudes via two approaches. The first one is utilizing traditional factorizations on poles, while the second one is exploiting $2$-splits.

\begin{figure}
  \centering
   \includegraphics[width=4cm]{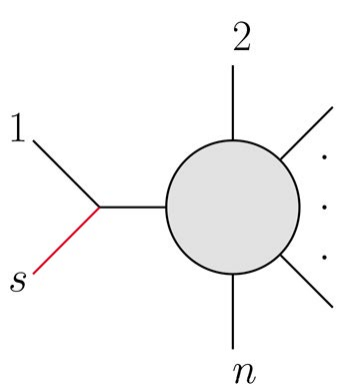}
   ~~~~~~~~~~~~~~~~~~~~
   \includegraphics[width=4cm]{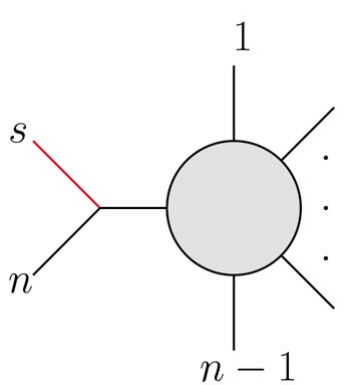} \\
  \caption{Diagrams with divergent propagators in the single-soft limit, where the soft particle represented in red is attached to the cubic vertex in each diagram.}\label{pole-part}
\end{figure}

We begin with the first method. In the soft limit, the leading contributions obviously arise from divergent propagators for $2$-point channels, as shown in Fig.\ref{pole-part}. On each pole $s_{s1}=0$ or $s_{ns}=0$, the full amplitude factorizes into two on-shell sub-amplitudes, one is the $3$-point amplitude expressed as a trivial constant, while another one is the $n$-point amplitude ${\cal A}_{{\rm Tr}(\phi^3)}(1,\cdots,n)$. Thus the soft theorem reads
\bea
{\cal A}^{(0)}_{{\rm Tr}(\phi^3)}(1,\cdots,n,s)={1\over\tau}\,\Big({1\over \hat{s}_{s1}}+{1\over \hat{s}_{ns}}\Big)\,{\cal A}_{{\rm Tr}(\phi^3)}(1,\cdots,n)\,.~~\label{softtheo-phi3-leading}
\eea

\begin{figure}
  \centering
   \includegraphics[width=12cm]{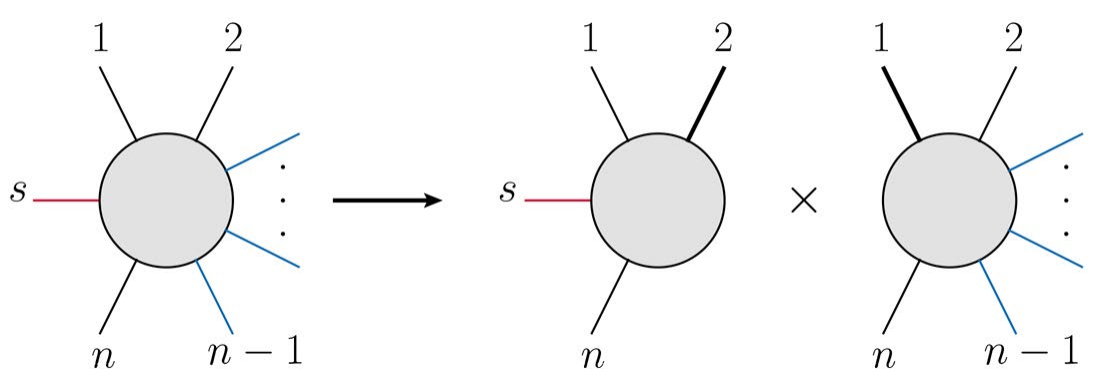}\\
   ~~~~~~~~\\
   ~~~~~~~~\\
   \includegraphics[width=12cm]{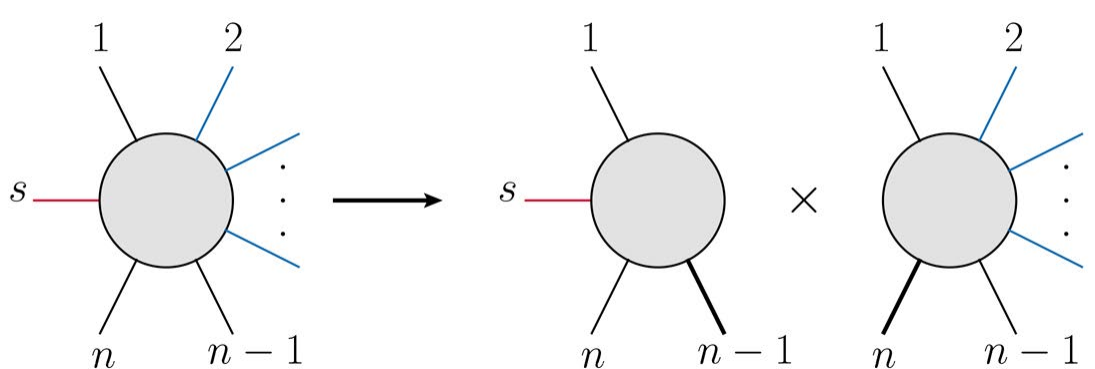} \\
  \caption{Two specific $2$-splits for ${\rm Tr}(\phi^3)$/YM amplitudes, where the set $A$ includes only one external leg $s$ represented in red, while legs in $B$ are represented in blue. The bold lines again represent off-shell particles.}\label{split-part}
\end{figure}

Now we show that the soft theorem \eref{softtheo-phi3-leading} can also be completely determined by considering $2$-splits. For the $n+1$-point amplitude ${\cal A}_{{\rm Tr}(\phi^3)}(1,\cdots,n,s)$, let us assign sets $A$ and $B$ as $A=\{s\}$, $B=\{3,\cdots,n-1\}$. Under such choice, the $2$-split in \eref{2split-phi3-special} is specialized into
\bea
{\cal A}_{{\rm Tr}(\phi^3)}(1,\cdots,n,s)\,\xrightarrow[B=\{3,\cdots,n-1\}]{k_s\cdot k_b=0}\,{\cal J}_{{\rm Tr}(\phi^3)}(n,s,1,2^*)\,\times\,{\cal J}_{{\rm Tr}(\phi^3)}(1^*,\cdots,n)\,,~~\label{2split-phi3-leading-1}
\eea
for $b\in B$, as illustrated in the first line of Fig.\ref{split-part}.
The key observation is, at the leading order, the soft limit $k_s\to 0$ automatically satisfies the kinematic condition $k_s\cdot k_b=0$, thus the above $2$-split naturally includes all information of leading soft behavior. Thus we can extract the soft theorem \eref{softtheo-phi3-leading} as follows.
As introduced in section \ref{sec-2split}, if we use the momentum conservation to eliminate $k^*_2$ in ${\cal J}_{{\rm Tr}(\phi^3)}(n,s,1,2^*)$,
the expression of this current is the same as the on-shell amplitude, which can be directly figured out by using Feynman rules,
\bea
{\cal J}_{{\rm Tr}(\phi^3)}(n,s,1,2^*)={1\over s_{s1}}+{1\over s_{ns}}\,.~~\label{phi3-JL}
\eea
Meanwhile, in the current ${\cal J}_{{\rm Tr}(\phi^3)}(1^*,\cdots,n)$, the momentum conservation forces the off-shell momenta $k^*_1$ to be
$k^*_1=k_1+k_s$. In the soft limit, the leading contribution of ${\cal J}_{{\rm Tr}(\phi^3)}(1^*,\cdots,n)$ is the on-shell amplitude ${\cal A}_{{\rm Tr}(\phi^3)}(1,\cdots,n)$, namely,
\bea
{\cal J}^{(0)}_{{\rm Tr}(\phi^3)}(1^*,\cdots,n)={\cal A}_{{\rm Tr}(\phi^3)}(1,\cdots,n)\,.~~\label{phi3-JR}
\eea
Plugging \eref{phi3-JL} and \eref{phi3-JR} into \eref{2split-phi3-leading-1}, one immediately reproduce the soft theorem \eref{softtheo-phi3-leading}.

One can also chose the set $B$ to be $B=\{2,\cdots,n-2\}$, without altering the choice of $A$,
to get the $2$-split
\bea
{\cal A}_{{\rm Tr}(\phi^3)}(1,\cdots,n,s)\,\xrightarrow[B=\{2,\cdots,n-2\}]{k_s\cdot k_b=0}\,{\cal J}_{{\rm Tr}(\phi^3)}(n-1^*,n,s,1)\,\times\,{\cal J}_{{\rm Tr}(\phi^3)}(1,\cdots,n^*)\,,~~\label{2split-phi3-leading-2}
\eea
as shown in the second line of Fig.\ref{split-part}.
Then the leading soft theorem can be extracted from the above $2$-split, via the analogous argument.

We have shown that the leading single-soft theorem can be obtained by employing factorizations on poles, or utilizing $2$-splits. For the single-soft ${\rm Tr}(\phi^3)$ case, as well as the single-soft YM case which will be discussed in the next section, both of two methods are effective. However, for double-soft NLSM case under consideration in section \ref{sec-NLSM}, since the leading soft behavior is not divergent, factorizations at on-shell propagators are not sufficient to fix the leading soft behavior. On the other hand, as will be seen, the leading double-soft theorem of NLSM amplitudes can also be determined by using $2$-splits.

\subsection{Sub-leading order}
\label{subsec-subleading-phi3}

In this subsection, we study the soft theorem of ${\rm Tr}(\phi^3)$ amplitudes at sub-leading order. We will show that, the sub-leading soft factor can be completely fixed by combining the behavior near divergent propagators, and $2$-splits in \eref{2split-phi3-leading-1} and \eref{2split-phi3-leading-2}. Indeed, the sub-leading soft theorem of ${\rm Tr}(\phi^3)$ amplitudes can also be derived by analyzing solely Feynman diagrams, this method is more effective for the ${\rm Tr}(\phi^3)$ case under consideration. However, the first method is more general since it can be straightforwardly generalized to higher orders (in the reduced lower-dimensional kinematic space), as well as YM and NLSM cases in subsequent sections. Therefore, we demonstrate the first method in detail, and only give a very brief discussion about the second one at the end of this subsection.

We can separate the full amplitude into two parts
\bea
{\cal A}_{{\rm Tr}(\phi^3)}(1,\cdots,n,s)=\Big[{1\over s_{s1}}\,{\cal J}_{{\rm Tr}(\phi^3)}(1^*,\cdots,n)+{1\over s_{ns}}\,{\cal J}_{{\rm Tr}(\phi^3)}(1,\cdots,n^*)\Big]\,+\,{\cal R}\,,~~\label{decom-phi3}
\eea
where the first part corresponds to diagrams in Fig.\ref{pole-part}, while the remaining part ${\cal R}$ corresponds to diagrams without any divergent propagator in the soft limit. In the above, ${\cal J}_{{\rm Tr}(\phi^3)}(1^*,\cdots,n)$ and ${\cal J}_{{\rm Tr}(\phi^3)}(1,\cdots,n^*)$ are two currents, carry off-shell momenta $k_1^*=k_1+k_s$ and $k_n^*=k_n+k_s$, respectively. The leading soft behavior purely arises from the first part, while the higher order soft behaviors receive contributions from both of two parts.
The ansatz \eref{decom-phi3} bears the strong similarity with that in \cite{Arkani-Hamed:2024fyd}, and it is worth to clarify the difference between them. In \cite{Arkani-Hamed:2024fyd} the coefficients of $1/s_{s1}$ and $1/s_{ns}$ are chosen to be residues on these poles, thus are exactly on-shell amplitudes. On the other hand, in \eref{decom-phi3} the coefficients of these propagators are determined by Feynman diagrams, thus are off-shell  Berends-Giele currents.

To get the sub-leading contribution of the first part in \eref{decom-phi3}, one should expand two off-shell currents ${\cal J}_{{\rm Tr}(\phi^3)}(1^*,\cdots,n)$ and ${\cal J}_{{\rm Tr}(\phi^3)}(1,\cdots,n^*)$ by $\tau$, as in \eref{expan-multi}. For the choice $A=\{s\}$, such expansions become
\bea
{\cal J}_{{\rm Tr}(\phi^3)}(1^*,\cdots,n)&=&{\cal A}_{{\rm Tr}(\phi^3)}(1,\cdots,n)+\sum_{m=1}^\infty\,{\tau^m\over m!}\,(\hat{k}_s\cdot\partial_{k_1})^m\,{\cal A}_{{\rm Tr}(\phi^3)}(1,\cdots,n)\,,\nn
{\cal J}_{{\rm Tr}(\phi^3)}(1,\cdots,n^*)&=&{\cal A}_{{\rm Tr}(\phi^3)}(1,\cdots,n)+\sum_{m=1}^\infty\,{\tau^m\over m!}\,(\hat{k}_s\cdot\partial_{k_n})^m\,{\cal A}_{{\rm Tr}(\phi^3)}(1,\cdots,n)\,.~~~\label{expan-phi3current}
\eea
From above expansions, we find the sub-leading contributions
\bea
{\cal J}^{(1)}_{{\rm Tr}(\phi^3)}(1^*,\cdots,n)&=&\hat{k}_s\cdot\partial_{k_1}\,{\cal A}_{{\rm Tr}(\phi^3)}(1,\cdots,n)\,,\nn
{\cal J}^{(1)}_{{\rm Tr}(\phi^3)}(1,\cdots,n^*)&=&\hat{k}_s\cdot\partial_{k_n}\,{\cal A}_{{\rm Tr}(\phi^3)}(1,\cdots,n)\,.~~\label{sub-phi3-part1}
\eea
Plugging them into the separation \eref{decom-phi3}, the full sub-leading soft behavior reads
\bea
{\cal A}^{(1)}_{{\rm Tr}(\phi^3)}(1,\cdots,n,s)=\Big[{1\over \hat{s}_{s1}}\,\hat{k}_s\cdot\partial_{k_1}+{1\over \hat{s}_{ns}}\,\hat{k}_s\cdot\partial_{k_n}\Big]\,{\cal A}_{{\rm Tr}(\phi^3)}(1,\cdots,n)\,+\,{\cal R}^{(0)}\,.~~\label{sub-phi3-step1}
\eea

To access the remaining unknown part ${\cal R}^{(0)}$, we exploit $2$-splits \eref{2split-phi3-leading-1} and \eref{2split-phi3-leading-2}, rather than using only one $2$-split as the method in \cite{Arkani-Hamed:2024fyd}.
One can express these two splittings in more concrete formulae, by substituting the explicit expression of $4$-point currents in \eref{phi3-JL}, and expansions of two currents ${\cal J}_{{\rm Tr}(\phi^3)}(1^*,\cdots,n)$ and ${\cal J}_{{\rm Tr}(\phi^3)}(1,\cdots,n^*)$ in \eref{expan-phi3current},
\bea
& &{\cal A}_{{\rm Tr}(\phi^3)}(1,\cdots,n,s)\,\xrightarrow[B=\{3,\cdots,n-1\}]{k_s\cdot k_b=0}\nn
& &~~~~~~~~~~~~~~~~{1\over\tau}\,\Big({1\over \hat{s}_{s1}}+{1\over \hat{s}_{ns}}\Big)\,\times\,\Big[{\cal A}_{{\rm Tr}(\phi^3)}(1,\cdots,n)+\sum_{m=1}^{\infty}\,{1\over m!}\,(\tau\,\hat{k}_s\cdot \partial_{k_1})^m\,{\cal A}_{{\rm Tr}(\phi^3)}(1,\cdots,n)\Big]\,,\nn
& &{\cal A}_{{\rm Tr}(\phi^3)}(1,\cdots,n,s)\,\xrightarrow[B=\{2,\cdots,n-2\}]{k_s\cdot k_b=0}\nn
& &~~~~~~~~~~~~~~~~{1\over\tau}\,\Big({1\over \hat{s}_{s1}}+{1\over \hat{s}_{ns}}\Big)\,\times\,\Big[{\cal A}_{{\rm Tr}(\phi^3)}(1,\cdots,n)+\sum_{m=1}^{\infty}\,{1\over m!}\,(\tau\,\hat{k}_s\cdot \partial_{k_n})^m\,{\cal A}_{{\rm Tr}(\phi^3)}(1,\cdots,n)\Big]\,.~~\label{subcontri-split}
\eea
Comparing \eref{subcontri-split} with \eref{sub-phi3-step1}, we find that on special loci in kinematic space, the ${\cal R}^{(0)}$ part behaves as
\bea
{\cal R}^{(0)}\,&\xrightarrow[B=\{3,\cdots,n-1\}]{k_s\cdot k_b=0}&\,{1\over \hat{s}_{ns}}\,\hat{k}_s\cdot\big(\partial_{k_1}-\partial_{k_n}\big)\,{\cal A}_{{\rm Tr}(\phi^3)}(1,\cdots,n)\,,\nn
{\cal R}^{(0)}\,&\xrightarrow[B=\{2,\cdots,n-2\}]{k_s\cdot k_b=0}&\,{1\over \hat{s}_{s1}}\,\hat{k}_s\cdot\big(\partial_{k_n}-\partial_{k_1}\big)\,{\cal A}_{{\rm Tr}(\phi^3)}(1,\cdots,n)\,.~~\label{R0-constrain}
\eea
The above two formulae are not equivalent to each other, but the full ${\cal R}^{(0)}$ should be an unique one.
To restore the desired unique ${\cal R}^{(0)}$ from above behaviors, let us consider the effect of acting the operator $\partial_{k_1}-\partial_{k_n}$ on ${\cal A}_{{\rm Tr}(\phi^3)}(1,\cdots,n)$. Each ${\rm Tr}(\phi^3)$ amplitude consists of solely scalar propagators. The operator $\partial_{k_1}-\partial_{k_n}$ annihilates any propagator containing both $k_1$ and $k_n$, thus only acts on propagators including one of $k_1$ and $k_n$. In other words, this operator only acts on propagators along the line $L_{1,n}$ which connects two external legs $1$ and $n$. Propagators along $L_{1,n}$ can be parameterized as $1/s_{1\cdots i}$ or $1/s_{j\cdots n}$, with
$2\leq i\leq n-2$ and $3\leq j\leq n-1$. It is direct to verify that, acting the operator $\partial_{k_1}-\partial_{k_n}$ leads to the same resulting object for above two alternative parameterizations,
\bea
\big(\partial_{k_1}-\partial_{k_n}\big)^\mu\,{\cal A}_{{\rm Tr}(\phi^3)}(1,\cdots,n)&=&\sum_{i=2}^{n-2}\,2k^\mu_{1\cdots i}\,\partial_{s_{1\cdots i}}\,{\cal A}_{{\rm Tr}(\phi^3)}(1,\cdots,n)\nn
&=&-\sum_{j=3}^{n-1}\,2k^\mu_{j\cdots n}\,\partial_{s_{j\cdots n}}\,{\cal A}_{{\rm Tr}(\phi^3)}(1,\cdots,n)\,,~~\label{deri-phi3}
\eea
where the equivalence between two lines are ensured by momentum conservation. When deriving the above result, a subtle point is $\partial_{k_1}^\mu s_{1\cdots i}=2k^\mu_{1\cdots i}$ rather than $=2k^\mu_{2\cdots i}$, since the definition of $s_{1\cdots i}$ includes $k_1^2$. Substituting \eref{deri-phi3} into \eref{R0-constrain}, and using the observation
\bea
& &\hat{k}_s\cdot k_{j\cdots n}\,\xrightarrow[B=\{3,\cdots,n-1\}]{k_s\cdot k_b=0}\,\hat{k}_s\cdot k_n\,,~~~~~~~~
\hat{k}_s\cdot k_{1\cdots i}\,\xrightarrow[B=\{2,\cdots,n-2\}]{k_s\cdot k_b=0}\,\hat{k}_s\cdot k_1\,,\nn
& &~~~~~~~~~~~~~~~~~~~~~~~~{\rm for}~2\leq i\leq n-2\,,~~3\leq j\leq n-1\,,~~\label{observe-k.k}
\eea
we obtain
\bea
{\cal R}^{(0)}\,\xrightarrow[B=\{3,\cdots,n-1\}~{\rm or}~B=\{2,\cdots,n-2\}]{k_s\cdot k_b=0}\,-\sum_{i=2}^{n-2}\,\partial_{s_{1\cdots i}}\,{\cal A}_{{\rm Tr}(\phi^3)}(1,\cdots,n)\,.
\eea
The above unique formula, which is valid for both cases $B=\{3,\cdots,n-1\}$ and $B=\{2,\cdots,n-2\}$, leads to the natural conjecture
\bea
{\cal R}^{(0)}\,=\,-\sum_{i=2}^{n-2}\,\partial_{s_{1\cdots i}}\,{\cal A}_{{\rm Tr}(\phi^3)}(1,\cdots,n)\,,~~~\label{R0}
\eea
which gives rise to the sub-leading soft theorem
\bea
{\cal A}^{(1)}_{{\rm Tr}(\phi^3)}(1,\cdots,n,s)=\Big[{1\over \hat{s}_{s1}}\,\hat{k}_s\cdot\partial_{k_1}+{1\over \hat{s}_{ns}}\,\hat{k}_s\cdot\partial_{k_n}-\sum_{i=2}^{n-2}\,\partial_{s_{1\cdots i}}\Big]\,{\cal A}_{{\rm Tr}(\phi^3)}(1,\cdots,n),~~~\label{soft-theo-phi3-sub}
\eea
where the ansatz \eref{sub-phi3-step1} has been used.

Logically, one can not conclude that the soft factor given in \eref{soft-theo-phi3-sub} is the complete answer in the full kinematic space, since it only satisfies Feynman rules for diagrams which contain the propagator $1/s_{ns}$ or $1/s_{s1}$, and offers proper $2$-splits on two special loci. However, one can argue the correctness of \eref{soft-theo-phi3-sub} by using the feature that the ${\cal R}^{(0)}$ part should not contain $1/s_{s1}$ or $1/s_{ns}$, due to the ansatz \eref{sub-phi3-step1}. The only possible source of potential missed terms in \eref{R0} is $k_s\cdot k_b=0$, with $b\in\{3,\cdots,n-2\}$ obtained by $\{2,\cdots,n-2\}\cap\{3,\cdots,n-1\}$, since ${\cal R}^{(0)}$ in \eref{R0} holds for $2$-splits with either $B=\{2,\cdots,n-2\}$ or $B=\{3,\cdots,n-1\}$. However, the ansatz \eref{sub-phi3-step1} indicates that each denominator in the ${\cal R}^{(0)}$ part is at the $\tau^0$ order, thus $k_s$ can never enter the numerator of ${\cal R}^{(0)}$, otherwise the corresponding contribution is at the $\tau^1$ order rather than $\tau^0$. This observation excludes the possibility $k_s\cdot k_b=0$.

As will be discussed in subsection \ref{subsec-check-phi3}, $-\partial_{s_{1\cdots i}}$ in \eref{R0} and \eref{soft-theo-phi3-sub} is a nice operator which makes the full sub-leading soft factor to be consistent with the momentum conservation law. When seeking higher order soft theorems in the next subsection, we expect the corresponding ${\cal R}^{m}$ to capture the similar consistency.

\begin{figure}
  \centering
   \includegraphics[width=8cm]{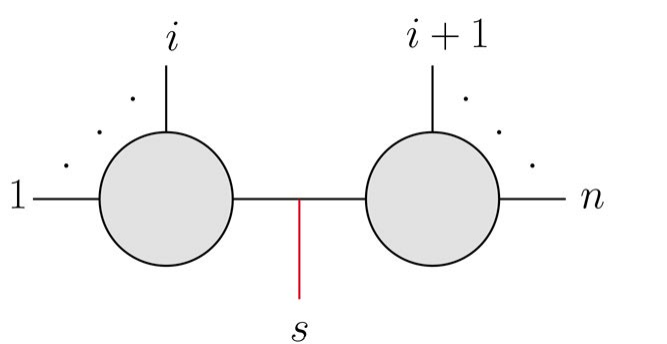}\\
  \caption{Diagrammatic interpretation of ${\cal R}_0$ in \eref{R0}, where the soft particle represented in red is attached to an internal line.}\label{inter-part}
\end{figure}

The soft sub-leading theorem \eref{soft-theo-phi3-sub} can also be derived by analyzing solely Feynman diagrams, since the ${\cal R}^{(0)}$ part has a straightforwardly graphic interpretation in Fig.\ref{inter-part}.
Acting $-\partial_{s_{1\cdots i}}$ turns $1/s_{1\cdots i}$ to $1/s_{1\cdots i}^2$, such doubled propagator $1/s_{1\cdots i}^2$ can be directly observed in Fig.\ref{inter-part}, by taking the limit $k_s\to 0$. Diagrams characterized by Fig.\ref{inter-part}, together with those characterized by Fig.\ref{pole-part} which correspond to the first part in \eref{sub-phi3-step1}, cover all allowed diagrams for a tree-level ${\rm Tr}(\phi^3)$
amplitude, thus ${\cal R}^{(0)}$ found in \eref{R0} completes the sub-leading soft behavior in \eref{sub-phi3-step1}.

%

\subsection{Higher order}
\label{subsec-higher-phi3}

Now we apply the analogous procedure to higher orders. we will find a general formula \eref{softtheo-phi3-high} of soft theorem with the soft factor given in \eref{softfac-phi3-high}, valid at $m^{\rm th}$ order for any $m\geq1$, which also covers the $m=1$ case discussed in the previous subsection \ref{subsec-subleading-phi3}. However, for $m\geq2$, the general formula only holds in the lower-dimensional kinematic space, defined by $k_s\cdot k_b=0$ with $b\in\{3,\cdots,n-2\}$.

Let us consider the soft behavior ${\cal A}^{(m)}_{{\rm Tr}(\phi^3)}(1,\cdots,n,s)$ with arbitrary positive integer $m$. By employing expansions \eref{expan-phi3current}, the separation \eref{sub-phi3-step1} can be rewritten as
\bea
{\cal A}^{(m)}_{{\rm Tr}(\phi^3)}(1,\cdots,n,s)={1\over m!}\,\Big[{\tau^{m-1}\over \hat{s}_{s1}}\,(\hat{k}_s\cdot\partial_{k_1})^m+{\tau^{m-1}\over \hat{s}_{ns}}\,(\hat{k}_s\cdot\partial_{k_n})^m\Big]\,{\cal A}_{{\rm Tr}(\phi^3)}(1,\cdots,n)\,+\,{\cal R}^{(m-1)}\,,~~\label{high-phi3-step1}
\eea
which describes the soft behavior at the $m^{\rm th}$ order. To detect ${\cal R}^{(m-1)}$, we compare $2$-splits in \eref{subcontri-split}
with the separation \eref{high-phi3-step1}, to obtain the generalization of \eref{R0-constrain}
\bea
{\cal R}^{(m-1)}\,&\xrightarrow[B=\{3,\cdots,n-1\}]{k_s\cdot k_b=0}&\,{1\over m!}\,{\tau^{m-1}\over \hat{s}_{ns}}\,\Big[(\hat{k}_s\cdot\partial_{k_1})^{m}-(\hat{k}_s\cdot\partial_{k_n})^{m}\Big]\,{\cal A}_{{\rm Tr}(\phi^3)}(1,\cdots,n)\,,\nn
{\cal R}^{(m-1)}\,&\xrightarrow[B=\{2,\cdots,n-2\}]{k_s\cdot k_b=0}&\,{1\over m!}\,{\tau^{m-1}\over \hat{s}_{s1}}\,\Big[(\hat{k}_s\cdot\partial_{k_n})^{m}-(\hat{k}_s\cdot\partial_{k_1})^{m}\Big]\,{\cal A}_{{\rm Tr}(\phi^3)}(1,\cdots,n)\,.~~\label{Rm-constrain}
\eea
By plugging observations \eref{deri-phi3} and \eref{observe-k.k}, as well as
\bea
(\hat{k}_s\cdot\partial_{k_1})^{m}-(\hat{k}_s\cdot\partial_{k_n})^{m}=\hat{k}_s\cdot(\partial_{k_1}-\partial_{k_n})\,
\Big(\sum_{q=0}^{m-1}\,(\hat{k}_s\cdot\partial_{k_1})^q
\,(\hat{k}_s\cdot\partial_{k_n})^{m-1-q}\Big)\,,
\eea
we find
\bea
{\cal R}^{(m-1)}\,& &\xrightarrow[B=\{3,\cdots,n-1\}~{\rm or}~B=\{2,\cdots,n-2\}]{k_s\cdot k_b=0}\nn
& &-{\tau^{m-1}\over m!}\,\Big(\sum_{q=0}^{m-1}\,(\hat{k}_s\cdot\partial_{k_1})^q
\,(\hat{k}_s\cdot\partial_{k_n})^{m-1-q}\Big)\,\Big(\sum_{i=2}^{n-2}\,\partial_{s_{1\cdots i}}\Big)\,{\cal A}_{{\rm Tr}(\phi^3)}(1,\cdots,n)\,,~~~~\label{to-Rm}
\eea
which leads to the conjecture
\bea
{\cal R}^{(m-1)}\,=\,-{\tau^{m-1}\over m!}\,\Big(\sum_{q=0}^{m-1}\,(\hat{k}_s\cdot\partial_{k_1})^q
\,(\hat{k}_s\cdot\partial_{k_n})^{m-1-q}\Big)\,\Big(\sum_{i=2}^{n-2}\,\partial_{s_{1\cdots i}}\Big)\,{\cal A}_{{\rm Tr}(\phi^3)}(1,\cdots,n)\,.~~\label{Rm}
\eea
As can be seen, the above manipulation only converts the operator $\hat{k}_s\cdot(\partial_{k_1}-\partial_{k_n})$ to $-\sum_{i=2}^{n-2}\,\partial_{s_{1\cdots i}}$, this effect is totally the same as in the previous subsection \ref{subsec-leading-phi3}.
Putting \eref{high-phi3-step1} and \eref{Rm} together, we finally get the soft theorem at the $m^{\rm th}$ order for any integer $m\geq 1$,
\bea
{\cal A}^{(m)}_{{\rm Tr}(\phi^3)}(1,\cdots,n,s)\,=\,\tau^{m-1}\,S^{(m)}_{{\rm Tr}(\phi^3)}\,{\cal A}^{(m)}_{{\rm Tr}(\phi^3)}(1,\cdots,n)\,,~~\label{softtheo-phi3-high}
\eea
where the operator $S^{(m)}_{{\rm Tr}(\phi^3)}$ is given as
\bea
S^{(m)}_{{\rm Tr}(\phi^3)}={1\over m!}\,\Big[{1\over \hat{s}_{s1}}\,(\hat{k}_s\cdot\partial_{k_1})^m+{1\over \hat{s}_{ns}}\,(\hat{k}_s\cdot\partial_{k_n})^m-
\Big(\sum_{q=0}^{m-1}\,(\hat{k}_s\cdot\partial_{k_1})^q
\,(\hat{k}_s\cdot\partial_{k_n})^{m-1-q}\Big)\,\Big(\sum_{i=2}^{n-2}\,\partial_{s_{1\cdots i}}\Big)\Big]\,.~~~\label{softfac-phi3-high}
\eea

For the above general soft theorem in \eref{softtheo-phi3-high} and \eref{softfac-phi3-high}, with arbitrary value of $m$, the diagrammatic understanding is not as transparently as in the case $m=1$, thus it is not easy to derive \eref{softtheo-phi3-high} and \eref{softfac-phi3-high} via the second method described at the end of the previous subsection \ref{subsec-subleading-phi3}, but the first method in subsection \ref{subsec-subleading-phi3} is still effective. However, such general formula only describes the complete soft behavior of the $m=1$ case, as discussed in subsection \ref{subsec-subleading-phi3}. For $m\geq2$ cases, such general soft theorem is not the complete one in the full kinematic space, as can be verified via the directly computation. Instead, it serves as the correct soft theorem in the subspace defined by $k_s\cdot k_b=0$ with $b\in\{3,\cdots,n-2\}$,
as indicated by $\{2,\cdots,n-2\}\cap\{3,\cdots,n-1\}$, since this general formula satisfies $2$-splits with either $B=\{2,\cdots,n-2\}$ or $B=\{3,\cdots,n-1\}$.
It is worth to compare the $m\geq2$ cases with $m=1$, and find the reason why $m=1$ is special. Firstly, in the $m=1$ case, we can compare $2$-splits with the ansatz \eref{sub-phi3-step1} to find the behaviors of ${\cal R}^{(0)}$ on special loci, then determine the full ${\cal R}^{(0)}$ by imposing the uniqueness. For the $m\geq2$ cases, the analogous procedure leads to the result which is totally the same as in the $m=1$ case, as pointed out below \eref{Rm}. In other words, the method in the previous subsection \ref{subsec-subleading-phi3} can not detect information beyond the $m=1$ order. Secondly, although the soft theorem with $m=1$ derived in subsection \ref{subsec-subleading-phi3} is also logically restricted in the subspace determined by $k_s\cdot k_b=0$, but as we have seen, one can use the ansatz \eref{sub-phi3-step1} to forbid any $\hat{k}_s\cdot k_b$ in ${\cal R}^{(0)}$, then argue the correctness of \eref{soft-theo-phi3-sub}. However, the similar argument does not make sense for $m\geq2$ cases, since $\hat{k}_s$ which is accompanied with $\tau$ must enter the numerators of ${\cal R}^{(m)}$ at higher orders. Hence, there is no reason to expect the general formula \eref{softtheo-phi3-high} to be valid beyond the lower-dimensional subspace. On the other hand, we have verified until $8$-points and $m=3$, the general formula \eref{softtheo-phi3-high} dose serve as the correct soft theorem in the subspace.
In the next subsection, we will show that the general formula \eref{softtheo-phi3-high} is consistent with momentum conservation.

\subsection{Consistency with momentum conservation}
\label{subsec-check-phi3}

In this subsection, we show that the conjectured general soft theorem \eref{softtheo-phi3-high} in the subspace, is consistent with momentum conservation.

As discussed in section \ref{sec-intro}, when saying the consistency with momentum conservation, we mean that the effect of acting the operator $S^{(m)}_{{\rm Tr}(\phi^3)}$ should be independent of re-expressing the $n$-point amplitude ${\cal A}^{(m)}_{{\rm Tr}(\phi^3)}(1,\cdots,n)$ in \eref{softtheo-phi3-high} via momentum conservation. Suppose we replace $k_i\in\{k_1,\cdots,k_n\}$ in the expression of ${\cal A}^{(m)}_{{\rm Tr}(\phi^3)}(1,\cdots,n)$ by the combination of remaining $n-1$ momenta in $\{k_1,\cdots,k_n\}$, the resulting object created by acting $S^{(m)}_{{\rm Tr}(\phi^3)}$ may have new expression, but must be equivalent to the previous one. To deal with the above requirement, we adopt the method in \cite{Cheung:2017ems}. For each $n$-point amplitude, let us introduce the momentum conservation operator
\bea
P_n=\sum_{i=1}^n\,K_i\cdot V\,,~~\label{defin-P}
\eea
where $V$ is a reference Lorentz vector, and $K_i$ are operators. When acting on a function ${\cal F}_n$ of $n$ external momenta, each $K_i$ takes the value of $k_i$ carried by ${\cal F}_n$. If a transmutation operator ${\cal T}$ turns an $n$-point amplitude ${\cal A}_n$ to another $n$-point one ${\cal A}'_n$, namely, ${\cal T}\,{\cal A}_n={\cal A}'_n$, then $P_n{\cal A}_n=0$ and $P_n{\cal A}'_n=0$ indicate the commutativity $[{\cal T}\,,\,P_n]\,{\cal A}_n=0$. Notice that we only require these two operators to be effectively commutable when acting on physical amplitudes ${\cal A}_n$. If such commutativity is satisfied, then the action of ${\cal T}$ is independent of re-parameterizing ${\cal A}_n$ via momentum conservation. For instance, the insertion operator ${\cal I}_{ikj}\equiv\partial_{\epsilon_k\cdot k_i}-\partial_{\epsilon_k\cdot k_j}$ defined in \cite{Cheung:2017ems} is obviously commutable with $P_n$. The independence of re-parameterization is also clear, supposing we replace $k_{i}$ in ${\cal A}_n$ by $-\sum_{j\neq i}k_j$, then the action of $-\partial_{\epsilon_k\cdot k_j}$ will compensate the action of $\partial_{\epsilon_k\cdot k_i}$, therefore the effect of acting ${\cal I}_{ikj}$ is unaltered.

For soft operators $S^{(m)}_{{\rm Tr}(\phi^3)}$ under consideration, the situation is slightly different, since this operator transmutes the $n$-point amplitude ${\cal A}_{{\rm Tr}(\phi^3)}(1,\cdots,n)$ to the $n+1$-point one ${\cal A}_{{\rm Tr}(\phi^3)}(1,\cdots,n,s)$. For this new case, we have
\bea
P_n\,{\cal A}_{{\rm Tr}(\phi^3)}(1,\cdots,n)=0\,,~~\label{P-np}
\eea
and
\bea
P_{n+1}\,{\cal A}_{{\rm Tr}(\phi^3)}(1,\cdots,n,s)=\Big(P_n+K_s\cdot V\Big)\,{\cal A}_{{\rm Tr}(\phi^3)}(1,\cdots,n,s)=0\,,~~\label{P-n+1p}
\eea
where the operator $P_n$ was defined in \eref{defin-P}, and the the definition of the operator $K_s$ is the same as other $K_i$.
Expanding ${\cal A}_{{\rm Tr}(\phi^3)}(1,\cdots,n,s)$ as
\bea
{\cal A}_{{\rm Tr}(\phi^3)}(1,\cdots,n,s)=\sum_{m=0}^\infty\,S^{(m)}_{{\rm Tr}(\phi^3)}\,{\cal A}_{{\rm Tr}(\phi^3)}(1,\cdots,n)\,,
\eea
and substituting \eref{P-np} and \eref{P-n+1p}, we get the following equation,
\bea
\Big(P_n+K_s\cdot V\Big)\,\Big(\sum_{m=0}^\infty\,\tau^{m-1}\,S^{(m)}_{{\rm Tr}(\phi^3)}\Big)\,{\cal A}_{{\rm Tr}(\phi^3)}(1,\cdots,n)
=\Big(\sum_{m=0}^\infty\,\tau^{m-1}\,S^{(m)}_{{\rm Tr}(\phi^3)}\Big)\,P_n\,{\cal A}_{{\rm Tr}(\phi^3)}(1,\cdots,n)\,.
\eea
Picking up terms at the $\tau^{m-1}$ order in the above equation, we find the commutation relation
\bea
[S^{(m)}_{{\rm Tr}(\phi^3)}\,,\,P_n]\,{\cal A}_{{\rm Tr}(\phi^3)}(1,\cdots,n)=\big(\hat{k}_s\cdot V\big)\,S^{(m-1)}_{{\rm Tr}(\phi^3)}\,{\cal A}_{{\rm Tr}(\phi^3)}(1,\cdots,n)\,.~~\label{commu-momen-conser}
\eea
As discussed before, if the above commutator is satisfied for any value of $m$, then the soft theorem at each order is independent of re-expressing ${\cal A}_{{\rm Tr}(\phi^3)}(1,\cdots,n)$ via momentum conservation, thus is consistent with momentum conservation.

Now we verify that soft operators in \eref{softfac-phi3-high} exactly satisfy the expected commutation relation \eref{commu-momen-conser}.
Using observations
\bea
& &[\hat{k}_s\cdot\partial_{k_a}\,,\,P_n]\,{\cal A}_{{\rm Tr}(\phi^3)}(1,\cdots,n)=\big(\hat{k}_s\cdot V\big)\,{\cal A}_{{\rm Tr}(\phi^3)}(1,\cdots,n)\,,\nn
& &~~~~~~~~~~~~~[\hat{k}_s\cdot\partial_{k_a}\,,\,\hat{k}_s\cdot V]=0\,,~~~~~{\rm for}~a=1,n\,,~~\label{veri-mom-obser0}
\eea
we find
\bea
[(\hat{k}_s\cdot\partial_{k_a})^m\,,\,P_n]\,{\cal A}_{{\rm Tr}(\phi^3)}(1,\cdots,n)=m\,\big(\hat{k}_s\cdot V\big)\,(\hat{k}_s\cdot\partial_{k_a})^{m-1}\,{\cal A}_{{\rm Tr}(\phi^3)}(1,\cdots,n)\,,~~\label{veri-mom-obser1}
\eea
and
\bea
& &[(\hat{k}_s\cdot\partial_{k_1})^q\,(\hat{k}_s\cdot\partial_{k_n})^{m-1-q}\,,\,P_n]\,{\cal A}_{{\rm Tr}(\phi^3)}(1,\cdots,n)\nn
&=&\big(\hat{k}_s\cdot V\big)\,\Big(q\,(\hat{k}_s\cdot\partial_{k_1})^{q-1}\,(\hat{k}_s\cdot\partial_{k_n})^{m-1-q}+(m-1-q)\,(\hat{k}_s\cdot\partial_{k_1})^q\,
(\hat{k}_s\cdot\partial_{k_n})^{m-2-q}
\Big)\,{\cal A}_{{\rm Tr}(\phi^3)}(1,\cdots,n)\,.~~\label{veri-mom-obser2}
\eea
Plugging \eref{veri-mom-obser1}, \eref{veri-mom-obser2} and \eref{softfac-phi3-high}, we then obtain
\bea
[S^{(m)}_{{\rm Tr}(\phi^3)}\,,\,P_n]\,{\cal A}_{{\rm Tr}(\phi^3)}(1,\cdots,n)={\hat{k}_s\cdot V\over m!}\,D\,{\cal A}_{{\rm Tr}(\phi^3)}(1,\cdots,n)\,,~~\label{commu-momen-step}
\eea
where the operator $D$ can be computed as
\bea
D&=&\Big[{m\over \hat{s}_{s1}}\,(\hat{k}_s\cdot\partial_{k_1})^{m-1}+{m\over \hat{s}_{s1}}\,(\hat{k}_s\cdot\partial_{k_n})^{m-1}\Big]\nn
& &-\Big(\sum_{i=2}^{n-2}\,\partial_{s_{1,\cdots i}}\Big)\,\Big[\sum_{q=0}^{m-1}\,\Big(q\,(\hat{k}_s\cdot\partial_{k_1})^{q-1}\,(\hat{k}_s\cdot\partial_{k_n})^{m-1-q}+(m-1-q)\,
(\hat{k}_s\cdot\partial_{k_1})^q\,(\hat{k}_s\cdot\partial_{k_n})^{m-2-q}
\Big)\Big]\nn
&=&\Big[{m\over \hat{s}_{s1}}\,(\hat{k}_s\cdot\partial_{k_1})^{m-1}+{m\over \hat{s}_{s1}}\,(\hat{k}_s\cdot\partial_{k_n})^{m-1}\Big]\nn
& &-\Big(\sum_{i=2}^{n-2}\,\partial_{s_{1,\cdots i}}\Big)\,\Big[\sum_{q'=0}^{m-2}\,(q'+1)\,(\hat{k}_s\cdot\partial_{k_1})^{q'}\,(\hat{k}_s\cdot\partial_{k_n})^{m-2-q'}+\sum_{q=0}^{m-2}\,(m-1-q)\,
(\hat{k}_s\cdot\partial_{k_1})^q\,(\hat{k}_s\cdot\partial_{k_n})^{m-2-q}\Big]\nn
&=&\Big[{m\over \hat{s}_{s1}}\,(\hat{k}_s\cdot\partial_{k_1})^{m-1}+{m\over s_{s1}}\,(\hat{k}_s\cdot\partial_{k_n})^{m-1}\Big]\,-\,\Big(\sum_{i=2}^{n-2}\,\partial_{s_{1,\cdots i}}\Big)\,\Big[\sum_{q=0}^{m-2}\,m\,(\hat{k}_s\cdot\partial_{k_1})^{q}\,(\hat{k}_s\cdot\partial_{k_n})^{m-2-q}\Big]\,.~~~\label{operator-D}
\eea
In the above, the second equality uses $q'=q-1$, and the third is obtained by relabeling $q'$ with $q$. Notice that when $q=0$, the operator $(\hat{k}_s\cdot\partial_{k_1})^{q-1}$
should be understood as $0$. The similar understanding holds for $(\hat{k}_s\cdot\partial_{k_n})^{m-2-q}$ when $q=m-1$. Comparing the operator $D$ in \eref{operator-D} with the definition of operator $S^{(m)}_{{\rm Tr}(\phi^3)}$ in \eref{softfac-phi3-high}, one immediately see $D=m!S^{(m-1)}_{{\rm Tr}(\phi^3)}$, thus \eref{commu-momen-step} is equivalent to the desired conclusion \eref{commu-momen-conser}.

\section{Single-soft theorems of YM amplitudes}
\label{sec-YM}

In this section, we investigate the single-soft theorems of tree-level YM amplitudes, via a similar procedure to the ${\rm Tr}(\phi^3)$ case.

As with the ${\rm Tr}(\phi^3)$ case, the leading soft theorem of YM amplitudes can be completely determined by utilizing either (i) Feynman diagrams with divergent propagators, or (ii) the $2$-split in \eref{2split-YM-special} with $A=\{s\}$, where $s$ denotes the soft external gluon. The detailed derivation is presented in subsection \ref{subsec-leading-YM}. The sub-leading soft theorem requires combining both ingredients, as shown in subsection \ref{subsec-subleading-YM}. To fully determine the complete sub-leading soft factor, we introduce a new technical approach that exploits momentum conservation. Unlike the ${\rm Tr}(\phi^3)$ case discussed in section \ref{subsec-subleading-phi3}, where the argument based on the order of $\tau$ suffice, the YM case requires additional verification. Therefore, to maintain methodological self-consistency, in subsection \ref{subsec-veri-YM-sub} we employ the more general 2-split \eref{2split-YM-general} to rigorously ensure the correctness of our result. This approach simultaneously yields the sub-leading soft behavior for one of the two currents in the 2-split \eref{2split-YM-general}.
In subsection \ref{subsec-higher-YM}, we extend the method from subsection \ref{subsec-subleading-YM} to higher orders, obtaining a universal representation of the higher order soft factor within a lower-dimensional subspace of the full kinematic space. Finally, in subsection \ref{subsec-check-YM}, we verify the resulting higher order soft theorem through three independent checks, which are
the gauge invariance preservation, the consistency with momentum conservation, as well as the proper reduction to the ${\rm Tr}(\phi^3)$ soft theorem under the transmutation operator proposed in \cite{Cheung:2017ems}.

\subsection{Leading order}
\label{subsec-leading-YM}

Similar as in \eref{decom-phi3}, we separate an $n+1$-point YM amplitude into two parts,
\bea
{\cal A}_{\rm YM}(1,\cdots,n,s)=\Big[{1\over s_{s1}}\,{\cal J}_{\rm YM}(s,1,I_1)\cdot{\cal J}_{\rm YM}(1^*,\cdots,n)+{1\over s_{ns}}\,{\cal J}_{\rm YM}(n,s,I_2)\cdot{\cal J}_{\rm YM}(1,\cdots,n^*)\Big]\,+\,{\cal R}\,.~~\label{decom-YM}
\eea
The first part, which is illustrated in Fig.\ref{pole-part}, contains divergent propagators in the soft limit, while the second part ${\cal R}$ does not. Four ${\cal J}^\mu_{\rm YM}$ in the above are YM vector currents, carry off-shell momenta $k_{I_1}=\sum_{i=2}^n k_i$, $k_{I_2}=\sum_{i=1}^{n-1} k_i$, $k^*_1=k_1+k_s$, $k^*_n=k_n+k_s$, respectively. The soft behavior at the leading order purely arises from the first part.
As in the ${\rm Tr}(\phi^3)$ case in section \ref{subsec-leading-phi3}, the leading soft theorem can be derived from either factorizations on poles, or $2$-splits.

We first focus on factorizations on poles. Using the Feynman rules for color-ordered amplitudes in Feynman gauge, two $3$-point currents ${\cal J}^\mu_{\rm YM}(s,1,I_1)$ and ${\cal J}^\mu_{\rm YM}(n,s,I_2)$ can be evaluated as
\bea
{\cal J}^\mu_{\rm YM}(s,1,I_1)&=&{1\over 2}\,\Big[(\epsilon_1\cdot\epsilon_s)\,(k_1-k_s)^\mu+\epsilon_s\cdot(k_{I_1}-k_1)\,\epsilon_1^\mu
+\epsilon_1\cdot(k_s-k_{I_1})\,\epsilon_s^\mu\Big]\,,\nn
{\cal J}^\mu_{\rm YM}(n,s,I_2)&=&{1\over 2}\,\Big[(\epsilon_n\cdot\epsilon_s)\,(k_s-k_n)^\mu+\epsilon_n\cdot(k_{I_2}-k_s)\,\epsilon_s^\mu
+\epsilon_s\cdot(k_n-k_{I_2})\,\epsilon_n^\mu\Big]\,.
\eea
For latter convenience, we employ momentum conservation, as well as on-shell conditions $\epsilon_1\cdot k_1=\epsilon_n\cdot k_n=\epsilon_s\cdot k_s=0$, to rewrite them as
\bea
{\cal J}^\mu_{\rm YM}(s,1,I_1)&=&{1\over 2}\,\Big[(\epsilon_1\cdot\epsilon_s)\,(-2\,\tau\,\hat{k}_s-k_{I_1})^\mu-2(\epsilon_s\cdot k_1)\,\epsilon_1^\mu
+2\tau(\epsilon_1\cdot \hat{k}_s)\,\epsilon_s^\mu\Big]\,,\nn
{\cal J}^\mu_{\rm YM}(n,s,I_2)&=&{1\over 2}\,\Big[(\epsilon_n\cdot\epsilon_s)\,(2\,\tau\,\hat{k}_s+k_{I_2})^\mu-2\tau(\epsilon_n\cdot \hat{k}_s)\,\epsilon_s^\mu
+2(\epsilon_s\cdot k_n)\,\epsilon_n^\mu\Big]\,,~~\label{J-3p-YM}
\eea
which have leading terms
\bea
{\cal J}^{\mu(0)}_{\rm YM}(s,1,I_1)&=&{1\over 2}\,\Big[-(\epsilon_1\cdot\epsilon_s)\,k_{I_1}^\mu-2(\epsilon_s\cdot k_1)\,\epsilon_1^\mu
\Big]\,,\nn
{\cal J}^{\mu(0)}_{\rm YM}(n,s,I_2)&=&{1\over 2}\,\Big[(\epsilon_n\cdot\epsilon_s)\,k_{I_2}^\mu
+2(\epsilon_s\cdot k_n)\,\epsilon_n^\mu\Big]\,.~~\label{J-3p-YM-leading}
\eea
Expressions in \eref{J-3p-YM} will also be used in the next subsection, since they also contain sub-leading terms.
Substituting leading terms in \eref{J-3p-YM-leading} into \eref{decom-YM}, and using
\bea
\epsilon_1\cdot{\cal J}^{(0)}_{\rm YM}(1^*,\cdots,n)=\epsilon_n\cdot{\cal J}^{(0)}_{\rm YM}(1,\cdots,n^*)={\cal A}_{\rm YM}(1,\cdots,n)\,,~~\label{J-YM-leading}
\eea
as well as the gauge invariance requirement\footnote{The Berends-Giele currents also satisfy "gauge invariance" in the sense of \eref{gauge-invariance-BG}, see in \cite{Berends:1987me,Wu:2021exa}}
\bea
k_{I_1}\cdot{\cal J}^{(0)}_{\rm YM}(1^*,\cdots,n)=k_{I_2}\cdot{\cal J}^{(0)}_{\rm YM}(1,\cdots,n^*)=0\,,~~\label{gauge-invariance-BG}
\eea
we arrive at the leading soft theorem
\bea
{\cal A}^{(0)}_{\rm YM}(1,\cdots,n,s)={1\over\tau}\,\Big({\epsilon_s\cdot k_n\over \hat{s}_{ns}}-{\epsilon_s\cdot k_1\over \hat{s}_{s1}}\Big)\,{\cal A}_{\rm YM}(1,\cdots,n)\,,~~\label{softtheo-YM-leading}
\eea
which is the same as the standard one \cite{Casali:2014xpa}.

Then we turn to the alternative method which utilizes $2$-splits. Similar as in the previous ${\rm Tr}(\phi^3)$ case, we assign $A$ and $B$ to be
$A=\{s\}$, $B=\{3,\cdots,n-1\}$ or $B=\{2,\cdots,n-2\}$. For the YM case, the $2$-splits are guaranteed not only by $k_a\cdot k_b=0$, but also by constraints on polarizations. We set
\bea
\epsilon_s\cdot k_b=0\,,~~~~\epsilon_i\cdot k_s=\epsilon_i\cdot\epsilon_s=0\,,~~~~{\rm for}~b\in B\,,~i\in\{1,\cdots,n\}\,,~~\label{condi-polar}
\eea
then the $2$-split \eref{2split-YM-special} is specialized as in Fig.\ref{split-part},
\bea
{\cal A}_{\rm YM}(1,\cdots,n,s)\,&\xrightarrow[B=\{3,\cdots,n-1\}]{k_s\cdot k_b=0\,,~\eref{condi-polar}}&\,{\cal J}_{{\rm YM}\oplus{\rm Tr}(\phi^3)}(n_\phi,s_g,1_\phi,2^*_\phi)\,\times\,\epsilon_1\cdot{\cal J}_{\rm YM}(1^*,\cdots,n)\,,\nn
{\cal A}_{\rm YM}(1,\cdots,n,s)\,&\xrightarrow[B=\{2,\cdots,n-2\}]{k_s\cdot k_b=0\,,~\eref{condi-polar}}&\,{\cal J}_{{\rm YM}\oplus{\rm Tr}(\phi^3)}(n-1^*_\phi,n_\phi,s_g,1_\phi)\,\times\,\epsilon_n\cdot{\cal J}_{\rm YM}(1,\cdots,n^*)\,.~~\label{2split-YM-soft}
\eea
Each $4$-point current carries three external scalars of ${\rm Tr}(\phi^3)$ theory which are labeled by the subscript $\phi$, and an external gluon $s_g$, as reviewed in section \ref{sec-2split}. We emphasize again, at the leading order the soft limit $k_s\to 0$ obeys the kinematic condition $k_s\cdot k_b=0$, thus each $2$-split in \eref{2split-YM-soft} incorporates all information of the leading soft behavior. To extract the leading soft behavior from these $2$-splits, one can eliminate $k^*_2$ in ${\cal J}_{{\rm YM}\oplus{\rm Tr}(\phi^3)}(n_\phi,s_g,1_\phi,2^*_\phi)$, and $k^*_{n-1}$ in ${\cal J}_{{\rm YM}\oplus{\rm Tr}(\phi^3)}(n-1^*_\phi,n_\phi,s_g,1_\phi)$, via momentum conservation, and use Feynman rules to evaluate these two currents as
\bea
{\cal J}_{{\rm YM}\oplus{\rm Tr}(\phi^3)}(n_\phi,s_g,1_\phi,2^*_\phi)={\cal J}_{{\rm YM}\oplus{\rm Tr}(\phi^3)}(n-1^*_\phi,n_\phi,s_g,1_\phi)
={\epsilon_s\cdot k_n\over s_{ns}}-{\epsilon_s\cdot k_1\over s_{s1}}\,.~~\label{J-YMphi-4p}
\eea
Substituting \eref{J-YMphi-4p} into \eref{2split-YM-soft}, and using the observation \eref{J-YM-leading}, the leading soft theorem \eref{softtheo-YM-leading} is reproduced.

\subsection{Sub-leading order}
\label{subsec-subleading-YM}

Now we turn to the sub-leading order. Similar as in section \ref{subsec-subleading-phi3} in the ${\rm Tr}(\phi^3)$ case, such sub-leading soft theorem of YM amplitudes can be derived by considering both the behavior near divergent propagators, and $2$-splits. On the other hand, the second method in section \ref{subsec-subleading-phi3} which exploiting solely Feynman diagrams is not effective in the current YM case.

According to the separation \eref{decom-YM}, the sub-leading soft behavior can be represented as
\bea
{\cal A}^{(1)}_{\rm YM}(1,\cdots,n,s)&=&{1\over\tau}\,\Big[{1\over \hat{s}_{s1}}\,{\cal J}^{(0)}_{\rm YM}(s,1,I_1)\cdot{\cal J}^{(1)}_{\rm YM}(1^*,\cdots,n)+{1\over \hat{s}_{ns}}\,{\cal J}^{(0)}_{\rm YM}(n,s,I_2)\cdot{\cal J}^{(1)}_{\rm YM}(1,\cdots,n^*)\nn
& &+{1\over \hat{s}_{s1}}\,{\cal J}^{(1)}_{\rm YM}(s,1,I_1)\cdot{\cal J}^{(0)}_{\rm YM}(1^*,\cdots,n)+{1\over \hat{s}_{ns}}\,{\cal J}^{(1)}_{\rm YM}(n,s,I_2)\cdot{\cal J}^{(0)}_{\rm YM}(1,\cdots,n^*)\Big]\nn
& &+{\cal R}^{(0)}\,.~~\label{YM-subleading}
\eea
Using expressions in \eref{J-3p-YM}, the sub-leading parts of ${\cal J}^\mu_{\rm YM}(s,1,I_1)$ and ${\cal J}^\mu_{\rm YM}(n,s,I_2)$ can be figured out as
\bea
{\cal J}^{\mu(1)}_{\rm YM}(s,1,I_1)&=&\tau\,\Big[-(\epsilon_1\cdot\epsilon_s)\,\hat{k}_s^\mu
+(\epsilon_1\cdot \hat{k}_s)\,\epsilon_s^\mu\Big]\,,\nn
{\cal J}^{\mu(1)}_{\rm YM}(n,s,I_2)&=&\tau\,\Big[(\epsilon_n\cdot\epsilon_s)\,\hat{k}_s^\mu-(\epsilon_n\cdot \hat{k}_s)\,\epsilon_s^\mu
\Big]\,,~~\label{J-3p-YM-sub}
\eea
Meanwhile, one can expand ${\cal J}^\mu_{\rm YM}(1^*,\cdots,n)$ and ${\cal J}^\mu_{\rm YM}(1,\cdots,n^*)$ as
\bea
\epsilon_1\cdot{\cal J}^\mu_{\rm YM}(1^*,\cdots,n)&=&{\cal A}_{\rm YM}(1,\cdots,n)+\sum_{m=1}^\infty\,{\tau^m\over m!}\,(\hat{k}_s\cdot\partial k_1)^m\,
{\cal A}_{\rm YM}(1,\cdots,n)\,,\nn
\epsilon_n\cdot{\cal J}^\mu_{\rm YM}(1,\cdots,n^*)&=&{\cal A}_{\rm YM}(1,\cdots,n)+\sum_{m=1}^\infty\,{\tau^m\over m!}\,(\hat{k}_s\cdot\partial k_n)^m\,
{\cal A}_{\rm YM}(1,\cdots,n)\,,~~\label{expan-J-np-YM}
\eea
therefore
\bea
\epsilon_1\cdot{\cal J}^{\mu(1)}_{\rm YM}(1^*,\cdots,n)&=&\tau\,(\hat{k}_s\cdot\partial k_1)\,
{\cal A}_{\rm YM}(1,\cdots,n)\,,\nn
\epsilon_1\cdot{\cal J}^{\mu(1)}_{\rm YM}(1,\cdots,n^*)&=&\tau\,(\hat{k}_s\cdot\partial k_n)\,
{\cal A}_{\rm YM}(1,\cdots,n)\,.~~\label{J-YM-sub}
\eea
Plugging \eref{J-3p-YM-leading}, \eref{J-YM-leading}, \eref{J-3p-YM-sub} and \eref{J-YM-sub} into \eref{YM-subleading}, we find
\bea
{\cal A}^{(1)}_{\rm YM}(1,\cdots,n,s)&=&\Big[{\epsilon_s\cdot k_n\over \hat{s}_{ns}}\,\hat{k}_s\cdot\partial_{k_n}-{\epsilon_s\cdot k_1\over \hat{s}_{s1}}\,\hat{k}_s\cdot\partial_{k_1}-{\epsilon_n\cdot \hat{f}_s\cdot\partial_{\epsilon_n}\over \hat{s}_{ns}}+{\epsilon_1\cdot \hat{f}_s\cdot\partial_{\epsilon_1}\over \hat{s}_{s1}}\Big]\,{\cal A}_{\rm YM}(1,\cdots,n)\nn
& &+{\cal R}^{(0)}\,,~~\label{sub-YM-step1}
\eea
where the strength tensor $\hat{f}_s^{\mu\nu}$ is defined as $\hat{f}_s^{\mu\nu}\equiv \hat{k}_s^\mu\epsilon_s^\nu-\epsilon_s^\mu \hat{k}_s^\nu$, and the linearity of each polarization is used to obtain operators $\epsilon_n\cdot \hat{f}_s\cdot\partial_{\epsilon_n}$ and $\epsilon_1\cdot \hat{f}_s\cdot\partial_{\epsilon_1}$.

Then we need to fix the ${\cal R}^{(0)}$ part by exploiting $2$-splits in \eref{2split-YM-soft}. Substituting \eref{J-YMphi-4p} and
\eref{J-YM-sub} into \eref{2split-YM-soft}, we obtain
\bea
{\cal A}^{(1)}_{\rm YM}(1,\cdots,n,s)\,&\xrightarrow[B=\{3,\cdots,n-1\}]{k_s\cdot k_b=0\,,~\eref{condi-polar}}&\,\Big({\epsilon_s\cdot k_n\over \hat{s}_{ns}}-{\epsilon_s\cdot k_1\over \hat{s}_{s1}}\Big)\,\times\,\hat{k}_s\cdot\partial_{k_1}\,{\cal A}_{\rm YM}(1,\cdots,n)\,,\nn
{\cal A}^{(1)}_{\rm YM}(1,\cdots,n,s)\,&\xrightarrow[B=\{2,\cdots,n-2\}]{k_s\cdot k_b=0\,,~\eref{condi-polar}}&\,\Big({\epsilon_s\cdot k_n\over \hat{s}_{ns}}-{\epsilon_s\cdot k_1\over \hat{s}_{s1}}\Big)\,\times\,\hat{k}_s\cdot\partial_{k_n}\,{\cal A}_{\rm YM}(1,\cdots,n)\,.~~\label{2split-YM-sub}
\eea
Comparing \eref{2split-YM-sub} with \eref{sub-YM-step1}, we observe the following behavior of ${\cal R}^{(0)}$ on special loci in kinematic space,
\bea
{\cal R}^{(0)}\,&\xrightarrow[B=\{3,\cdots,n-1\}]{k_s\cdot k_b=0\,,~\eref{condi-polar}}&\,{\epsilon_s\cdot k_n\over \hat{s}_{ns}}\,\hat{k}_s\cdot(\partial_{k_1}-\partial_{k_n})\,{\cal A}_{\rm YM}(1,\cdots,n)\,,\nn
{\cal R}^{(0)}\,&\xrightarrow[B=\{2,\cdots,n-2\}]{k_s\cdot k_b=0\,,~\eref{condi-polar}}&\,{\epsilon_s\cdot k_1\over \hat{s}_{s1}}\,\hat{k}_s\cdot(\partial_{k_1}-\partial_{k_n})\,{\cal A}_{\rm YM}(1,\cdots,n)\,.~~\label{R0-constrain-YM}
\eea
Notice that the operators $\epsilon_n\cdot \hat{f}_s\cdot\partial_{\epsilon_n}$ and $\epsilon_1\cdot \hat{f}_s\cdot\partial_{\epsilon_1}$ vanish due to the constraint \eref{condi-polar}.

Since YM amplitudes consists of not only propagators, one can not bootstrap ${\cal R}^{(0)}$ along the line in section \ref{subsec-higher-phi3}.
To proceed, we recall the discussion in section \ref{subsec-check-phi3} about the consistency with momentum conservation. To keep such consistency, soft operators should obey some constraints such as commutators \eref{commu-momen-conser}. It is natural to expect that the consistency with momentum conservation also imposes constraints to soft factors in the YM case, thus we can try to figure out ${\cal R}^{(0)}$ by considering the implication of momentum conservation.

Let us encode the vector $(\partial_{k_1}-\partial_{k_n})^\mu\,{\cal A}_{\rm YM}(1,\cdots,n)$ as
\bea
(\partial_{k_1}-\partial_{k_n})^\mu\,{\cal A}_{\rm YM}(1,\cdots,n)=\sum_i\,{\cal V}_i^\mu\,A^d_i\,,~~\label{defin-effc-partial}
\eea
where each ${\cal V}^\mu_i$ is a linear combination of external momenta and polarizations carried by ${\cal A}_{\rm YM}(1,\cdots,n)$, and the super script $d$ means each $A^d_i$ is obtained by taking derivative. In the first line of \eref{R0-constrain-YM}, only $k_1$, $k_2$ and $k_n$ involved in the vector ${\cal V}^\mu_i$ survive, due to conditions $k_s\cdot k_b=0$ and \eref{condi-polar}. Therefore, we can re-express the first line of \eref{R0-constrain-YM} as
\bea
{\cal R}^{(0)}\,&\xrightarrow[B=\{3,\cdots,n-1\}]{k_s\cdot k_b=0\,,~\eref{condi-polar}}&\,{\epsilon_s\cdot k_n\over \hat{s}_{ns}}\,\sum_i\,\hat{k}_s\cdot(c_{i1}\,k_1+c_{i2}\,k_2+c_{in}\,k_n)\,A^d_i\,.
\eea
The momentum conservation indicates, one can always modify the above form as
\bea
{\cal R}^{(0)}\,\xrightarrow[B=\{3,\cdots,n-1\}]{k_s\cdot k_b=0\,,~\eref{condi-polar}}&&\,{\epsilon_s\cdot k_n\over \hat{s}_{ns}}\,\sum_i\,\hat{k}_s\cdot\Big(c_{i1}\,k_1+c_{i2}\,k_2+c_{in}\,k_n+\W c_i\,\sum_{j=1}^n\,k_j\Big)\,A^d_i\nn
=&&\,{\epsilon_s\cdot k_n\over \hat{s}_{ns}}\,\sum_i\,\hat{k}_s\cdot\Big(c_{i1}\,k_1+c_{i2}\,k_2+c_{in}\,k_n+\W c_i\,(k_1+k_2+k_n)\Big)\,A^d_i\,,
\eea
with arbitrary complex numbers $\W c_i$, where the second equality uses $k_s\cdot k_b=0$. Choosing $\W c_i$ to be $\W c_i=-c_{i1}$ and $\W c_i=-c_{i2}$, we get
\bea
{\cal R}_1^{(0)}\,\xrightarrow[B=\{3,\cdots,n-1\}]{k_s\cdot k_b=0\,,~\eref{condi-polar}}\,{\epsilon_s\cdot k_n\over \hat{s}_{ns}}\,\sum_i\,\hat{k}_s\cdot\Big((c_{i2}-c_{i1})\,k_2+(c_{in}-c_{i1})\,k_n\Big)\,A^d_i\,,
\eea
and
\bea
{\cal R}_2^{(0)}\,\xrightarrow[B=\{3,\cdots,n-1\}]{k_s\cdot k_b=0\,,~\eref{condi-polar}}\,{\epsilon_s\cdot k_n\over \hat{s}_{ns}}\,\sum_i\,\hat{k}_s\cdot\Big((c_{i1}-c_{i2})\,k_1+(c_{in}-c_{i2})\,k_n\Big)\,A^d_i\,,
\eea
respectively. Of course, ${\cal R}_1^{(0)}$ should be equivalent to ${\cal R}_2^{(0)}$. The key observation is, ${\cal R}_1^{(0)}$ contains two parts, one has a pole $\hat{s}_{ns}= 0$, while another one does not have this pole since $\hat{k}_s\cdot k_n$ cancels $\hat{s}_{ns}$. The analogous observation holds for ${\cal R}_2^{(0)}$. Terms with or without pole should be identified individually, thus the following relation must hold, hence
\bea
(\hat{k}_s\cdot k_2)\,\sum_i\,(c_{i2}-c_{i1})\,A^d_i=(\hat{k}_s\cdot k_1)\,\sum_i\,(c_{i1}-c_{i2})\,A^d_i\,,
~~~~\sum_i\,(c_{in}-c_{i1})\,A^d_i=\sum_i\,(c_{in}-c_{i2})\,A^d_i\,.~~\label{equa-c}
\eea
The second equation in the above indicates,
\bea
\sum_i\,c_{i1}\,A^d_i=\sum_i\,c_{i2}\,A^d_i\,,~~\label{solu-c}
\eea
then the first equation is satisfied automatically.
Consequently, we have
\bea
{\cal R}^{(0)}\,\xrightarrow[B=\{3,\cdots,n-1\}]{k_s\cdot k_b=0\,,~\eref{condi-polar}}\,(\epsilon_s\cdot k_n)\,\sum_i\,{c_{in}-c_{i1}\over 2}\,A^d_i\,.~~\label{R0I-YM}
\eea
The analogous manipulation leads to
\bea
{\cal R}^{(0)}\,\xrightarrow[B=\{2,\cdots,n-2\}]{k_s\cdot k_b=0\,,~\eref{condi-polar}}\,(\epsilon_s\cdot k_1)\,\sum_i\,{c_{i1}-c_{in}\over2}\,A^d_i\,.~~\label{R0II-YM}
\eea
Similar as in section \ref{subsec-subleading-phi3}, we should impose the uniqueness to two different expressions of ${\cal R}^{(0)}$ in \eref{R0I-YM} and \eref{R0II-YM}. To get an unique ${\cal R}^{(0)}$, we notice that the solution \eref{solu-c}, together with the constraint \eref{condi-polar} on polarizations,
also indicate
\bea
\sum_i\,(\epsilon_s\cdot{\cal V}_i)\,A^d_i\,\xrightarrow[B=\{3,\cdots,n-1\}]{\eref{condi-polar}}\,\sum_i\,(c_{in}-c_{i1})\,(\epsilon_s\cdot k_n)\,A^d_i\,.~~\label{es-dot-kn}
\eea
The above behavior of $\epsilon_s\cdot{\cal V}_i$, together with the definition of ${\cal V}_i$ and $A^d_i$ in \eref{defin-effc-partial}, allow us to rewrite \eref{R0I-YM} as
\bea
{\cal R}^{(0)}\,\xrightarrow[B=\{3,\cdots,n-1\}]{k_s\cdot k_b=0\,,~\eref{condi-polar}}\,{1\over 2}\,\epsilon_s\cdot(\partial_{k_1}-\partial_{k_n})\,{\cal A}_{\rm YM}(1,\cdots,n)\,.~~\label{R0I-uni-YM}
\eea
The similar argument also turns \eref{R0II-YM} to
\bea
{\cal R}^{(0)}\,\xrightarrow[B=\{2,\cdots,n-2\}]{k_s\cdot k_b=0\,,~\eref{condi-polar}}\,{1\over 2}\,\epsilon_s\cdot(\partial_{k_1}-\partial_{k_n})\,{\cal A}_{\rm YM}(1,\cdots,n)\,,
\eea
which is exactly the same as in \eref{R0I-uni-YM}.
Thus, it is natural to conjecture
\bea
{\cal R}^{(0)}\,=\,{1\over 2}\,\epsilon_s\cdot(\partial_{k_1}-\partial_{k_n})\,{\cal A}_{\rm YM}(1,\cdots,n)\,.~~\label{R0-result-YM}
\eea
The above ${\cal R}^{(0)}$ turns \eref{sub-YM-step1} to
\bea
{\cal A}^{(1)}_{\rm YM}(1,\cdots,n,s)\,=\,S^{(1)}_{\rm YM}\,{\cal A}_{\rm YM}(1,\cdots,n)\,,~~\label{softtheo-YM-sub}
\eea
where
\bea
S^{(1)}_{\rm YM}=-{k_n\cdot \hat{f}_s\cdot\partial_{k_n}\over \hat{s}_{ns}}+{k_1\cdot \hat{f}_s\cdot\partial_{k_1}\over \hat{s}_{s1}}
-{\epsilon_n\cdot \hat{f}_s\cdot\partial_{\epsilon_n}\over \hat{s}_{ns}}+{\epsilon_1\cdot \hat{f}_s\cdot\partial_{\epsilon_1}\over \hat{s}_{s1}}\,.~~\label{softfac-YM-sub}
\eea

As shown in \cite{Zhou:2022orv,Du:2024dwm}, the soft factor \eref{softfac-YM-sub} is equivalent to the standard one
\bea
S^{(1)}_{\rm YM}={\epsilon_s\cdot J_n\cdot \hat{k}_s\over \hat{s}_{ns}}-{\epsilon_s\cdot J_1\cdot \hat{k}_s\over \hat{s}_{s1}}\,,
\eea
where each $J_i$ is understood as the angular momentum operator for the external particle $i$ \cite{Casali:2014xpa}. Thus the soft theorem \eref{softtheo-YM-sub} with  the soft factor given in \eref{softfac-YM-sub}
does serve as the correct sub-leading soft theorem of YM amplitudes. However, to make our method to be self contained, we should answer whether we can known the correctness of \eref{softtheo-YM-sub} without knowing the standard result. One straightforward evidence is the gauge invariance, the soft factor in \eref{softfac-YM-sub} automatically vanishes under the replacement $\epsilon_s\to \hat{k}_s$, due to the definition of $\hat{f}_s^{\mu\nu}$. Indeed, the ${\cal R}^{(0)}$ part \eref{R0-result-YM} can also be conjectured by imposing the gauge invariance to the ansatz \eref{sub-YM-step1}. However, the gauge invariance can not stop us from adding new operators to $S^{(1)}_{\rm YM}$ which are automatically gauge invariant, for instance the operators also contain $\hat{f}_s^{\mu\nu}$. Furthermore, as discussed in section \ref{sec-intro}, our purpose is to seek a general method which is also valid for theories without gauge invariance but exhibit $2$-split. Thus, we need a more general way to argue the correctness of \eref{softtheo-YM-sub}, such general way will be shown in the next subsection.

\subsection{Verification via $2$-splits}
\label{subsec-veri-YM-sub}

In this subsection, we suggest a method based on the more general $2$-split \eref{2split-YM-general}, to verify the correctness of conjectured sub-leading soft theorem of YM amplitudes. This method can be directly applied to other theories satisfying $2$-splits, such as NLSM which will be studied in the next section.
As a by product, this method also yields the sub-leading soft behavior of resulting currents in general $2$-splits under consideration, which is not straightforward to be evaluated via the definition.

Similar as in the ${\rm Tr}(\phi^3)$ case studied in section \ref{subsec-subleading-phi3}, logically the soft theorem \eref{softtheo-YM-sub} is restricted in the subspace defined by $k_s\cdot k_b=\epsilon_s\cdot k_b=k_s\cdot\epsilon_b=\epsilon_s\cdot\epsilon_b=0$, with $b\in\{3,\cdots,n-2\}$. In section \ref{subsec-subleading-phi3}, we argued the correctness of sub-leading soft theorem \eref{soft-theo-phi3-sub} in the full kinematic space, by showing the absence of $k_s$ in numerators of ${\cal R}^{(0)}$, due to the ansatz \eref{sub-phi3-step1}, as well as the fact that the sub-leading order is the $\tau^0$ order. For the YM case, this argument still makes sense, thus ${\cal R}^{(0)}$ in \eref{sub-YM-step1} can not contain terms like $k_s\cdot V$ where $V$ denotes the arbitrary Lorentz vector. However, this argument can not exclude the additional $\epsilon_s\cdot V$, where $V$ is the combination of $k_b$ and $\epsilon_b$ with $b\in\{3,\cdots,n-2\}$, since such $\epsilon_s\cdot V$ can not be detected in the subspace.
To completely show that \eref{softtheo-YM-sub} is the right answer, we exploit more general $2$-splits in \ref{2split-YM-general}. Let us choose $A=\{s\}$, $B=\{2,\cdots,n-1\}\setminus k$ with $k\neq1,2,n-1,n$. Then the $n+1$-point YM amplitude factorizes as in \eref{2split-YM-general},
\bea
{\cal A}_{\rm YM}(1,\cdots,n,s)\,&\xrightarrow[B=\{2,\cdots,n-1\}\setminus k]{k_s\cdot k_b=0\,,~\eref{condi-polar}}&\,{\cal J}_{{\rm YM}\oplus{\rm Tr}(\phi^3)}(n_\phi,s_g,1_\phi,\kappa_\phi)\,\times\,\epsilon_1\cdot{\cal J}_{\rm YM}(1,\cdots,\kappa',\cdots,n)\,,
\eea
where the position of off-shell legs $\kappa$ and $\kappa'$ in color orderings are inherited from $k$. Here the $4$-point current ${\cal J}_{{\rm YM}\oplus{\rm Tr}(\phi^3)}(n_\phi,s_g,1_\phi,\kappa_\phi)$ is the same as that in \eref{J-YMphi-4p}. In the soft limit, the sub-leading contribution of the above $2$-split reads
\bea
{\cal A}^{(1)}_{\rm YM}(1,\cdots,n,s)\,\xrightarrow[B=\{2,\cdots,n-1\}\setminus k]{k_s\cdot k_b=0\,,~\eref{condi-polar}}&&\,{\cal J}^{(1)}_{{\rm YM}\oplus{\rm Tr}(\phi^3)}(n_\phi,s_g,1_\phi,\kappa_\phi)\,\times\,\epsilon_1\cdot{\cal J}^{(0)}_{\rm YM}(1,\cdots,\kappa',\cdots,n)\nn
+&&\,{\cal J}^{(0)}_{{\rm YM}\oplus{\rm Tr}(\phi^3)}(n_\phi,s_g,1_\phi,\kappa_\phi)\,\times\,\epsilon_1\cdot{\cal J}^{(1)}_{\rm YM}(1,\cdots,\kappa',\cdots,n)\,,~~\label{veri-YM-sub-1}
\eea
Using the expression of $4$-point current in \eref{J-YMphi-4p}, we see that ${\cal A}^{(1)}_{\rm YM}(1,\cdots,n,s)$ only receives the contribution from the second line of \eref{veri-YM-sub-1}, and can be written more concretely as
\bea
{\cal A}^{(1)}_{\rm YM}(1,\cdots,n,s)\,\xrightarrow[B=\{2,\cdots,n-1\}\setminus k]{k_s\cdot k_b=0\,,~\eref{condi-polar}}&&\,\Big({\epsilon_s\cdot k_n\over s_{ns}}-{\epsilon_s\cdot k_1\over s_{s1}}\Big)\,\times\,\epsilon_1\cdot{\cal J}^{(1)}_{\rm YM}(1,\cdots,\kappa',\cdots,n)\,.~~\label{veri-YM-sub-2}
\eea
This is a condition which must be satisfied by the sub-leading soft behavior ${\cal A}^{(1)}_{\rm YM}(1,\cdots,n,s)$, for any choice of $k$. Choosing $k$ in turn through the set $\{3,\cdots,n-2\}$, one can detect any potential $\epsilon_s\cdot V$. More explicitly, suppose $\epsilon_s\cdot k_k$ or $\epsilon_s\cdot \epsilon_k$ is included in $\epsilon_s\cdot V$, such terms should occur in \eref{veri-YM-sub-2}, since $k_k$ and $\epsilon_k$ are not constrained by $k_a\cdot k_b=0$ and \eref{condi-polar} with $B=\{3,\cdots,n-1\}$ or $B=\{2,\cdots,n-2\}$. Thus one can fix the full ${\cal R}^{(0)}$ by imposing $2$-splits \eref{veri-YM-sub-2} with $k$ running through $\{3,\cdots,n-2\}$.

To do this, we first figure out the behavior of $S^{(1)}_{\rm YM}{\cal A}_{\rm YM}(1,\cdots,n)$, on the locus determined by $k_s\cdot k_b=0$ and \eref{condi-polar} with $B=\{2,\cdots,n-1\}\setminus k$. Clearly,
\bea
\{\hat{k}_s,\epsilon_s\}\,\cdot\,\partial_{k_n}\,{\cal A}_{\rm YM}(1,\cdots,n)&\,\xrightarrow[B=\{2,\cdots,n-1\}\setminus k]{k_s\cdot k_b=0\,,~\eref{condi-polar}}&\,\{\hat{k}_s,\epsilon_s\}\,\cdot\,\sum_{a=1,k,n}\,k_a\,\partial_{k_n\cdot k_a}\,{\cal A}_{\rm YM}(1,\cdots,n)\,,\nn
\{\hat{k}_s,\epsilon_s\}\,\cdot\,\partial_{k_1}\,{\cal A}_{\rm YM}(1,\cdots,n)&\,\xrightarrow[B=\{2,\cdots,n-1\}\setminus k]{k_s\cdot k_b=0\,,~\eref{condi-polar}}&\,\{\hat{k}_s,\epsilon_s\}\,\cdot\,\sum_{a=1,k,n}\,k_a\,\partial_{k_1\cdot k_a}\,{\cal A}_{\rm YM}(1,\cdots,n)\,,~~\label{ke-dot-remain}
\eea
where $\{\hat{k}_s,\epsilon_s\}$ means the Lorentz vector contracting with the remaining part can be either $\hat{k}_s$ or $\epsilon_s$.
The subtle point discussed below \eref{deri-phi3} also occurs in the above, hence we also inserted $\partial_{k_n\cdot k_n}$ and $\partial_{k_1\cdot k_1}$,
since these Lorentz invariants are contained in the definitions of Mandelstam variables thus should be considered when taking derivatives of $k_1$ and $k_n$.
A few algebra shows that the action of soft operator in \eref{softfac-YM-sub} leads to the factorization behavior
\bea
&&S^{(1)}_{\rm YM}\,{\cal A}_{\rm YM}(1,\cdots,n)\,\xrightarrow[B=\{2,\cdots,n-1\}\setminus k]{k_s\cdot k_b=0\,,~\eref{condi-polar}}\nn
&&~~~~~~~~~~~~\Big({\epsilon_s\cdot k_n\over s_{ns}}-{\epsilon_s\cdot k_1\over s_{s1}}\Big)\,\times\,\Big[\hat{k}_s\cdot k_1\,(\partial_{k_n\cdot k_1}-\partial_{k_n\cdot k_k})+\hat{k}_s\cdot k_n\,(\partial_{k_1\cdot k_n}-\partial_{k_1\cdot k_k})\Big]\,{\cal A}_{\rm YM}(1,\cdots,n)\,.~~\label{veri-YM-sub-3}
\eea
The detailed derivation of \eref{veri-YM-sub-3} is provided in Appendix \ref{appen}.
The factorization in \eref{veri-YM-sub-3} means, the soft theorem in \eref{softtheo-YM-sub} satisfies the required formula \eref{veri-YM-sub-2}.
Thus, if ${\cal A}^{(1)}_{\rm YM}(1,\cdots,n)$ contains additional terms proportional to $\epsilon_s\cdot k_k$ or $\epsilon_s\cdot\epsilon_k$, they
can only change the expression of $\epsilon_1\cdot{\cal J}^{(1)}_{\rm YM}(1,\cdots,\kappa',\cdots,n)$ in \eref{veri-YM-sub-3}. In other words, the undetected part should be turned into the factorized form
\bea
\Big({\epsilon_s\cdot k_n\over s_{ns}}-{\epsilon_s\cdot k_1\over s_{s1}}\Big)\,\times\,\cdots\,,
\eea
by transforming itself. It is obviously impossible, since $\epsilon_s\cdot k_n$ and $\epsilon_s\cdot k_1$ in missed terms can only arise from $\epsilon_s\cdot k_k=-\epsilon_s\cdot(k_1+k_n)$, but $\epsilon_s\cdot k_n$ and $\epsilon_s\cdot k_1$ created in this way must have the same coefficient. Consequently, for any choice of $k$, one can never add terms proportional to $\epsilon_s\cdot k_k$ or $\epsilon_s\cdot\epsilon_k$ into the sub-leading soft behavior, otherwise the factorization \eref{veri-YM-sub-2} will be violated. This observation ensures the correctness of the soft theorem given in \eref{softtheo-YM-sub} and \eref{softfac-YM-sub}.

Comparing \eref{veri-YM-sub-3} with \eref{veri-YM-sub-2}, one immediately see that the sub-leading soft behavior of the YM current ${\cal J}^{\mu(1)}_{\rm YM}(1,\cdots,\kappa',\cdots,n)$ is formally given as
\bea
\epsilon_1\cdot{\cal J}^{(1)}_{\rm YM}(1,\cdots,\kappa',\cdots,n)=\Big[\hat{k}_s\cdot k_1\,(\partial_{k_n\cdot k_1}-\partial_{k_n\cdot k_k})+\hat{k}_s\cdot k_n\,(\partial_{k_1\cdot k_n}-\partial_{k_1\cdot k_k})\Big]\,{\cal A}_{\rm YM}(1,\cdots,n)\,.~~\label{softtheo-YMcurrent}
\eea
Notice that we have not taken the momentum carried by any external leg to be soft. When saying soft behavior, we mean $k_s$ in $k_{\kappa'}=k_k+k_s$ is taken to be soft.
The above sub-leading soft behavior is not manifest from the definition of the current, however, we have verified it up to $6$-points.

\subsection{Higher order}
\label{subsec-higher-YM}

The technic in subsection \ref{subsec-subleading-YM} can be straightforwardly performed to higher orders. However, similar as in the ${\rm Tr}(\phi^3)$ case, for $m^{\rm th}$ order with $m\geq2$, this method only leads to the soft theorem in the lower-dimensional kinematic space, defined by $k_s\cdot k_b=0$ and \eref{condi-polar}, with $b\in\{3,\cdots,n-2\}$.

By plugging \eref{J-3p-YM} and \eref{expan-J-np-YM}, one can generalize the ansatz
\eref{sub-YM-step1} as
\bea
{\cal A}^{(m)}_{\rm YM}(1,\cdots,n,s)&=&{\tau^{m-1}\over m!}\,\Big[{\epsilon_s\cdot k_n\over \hat{s}_{ns}}\,(\hat{k}_s\cdot\partial_{k_n})^m-{\epsilon_s\cdot k_1\over \hat{s}_{s1}}\,(\hat{k}_s\cdot\partial_{k_1})^m\Big]\,{\cal A}_{\rm YM}(1,\cdots,n)\nn
& &-{\tau^{m-1}\over(m-1)!}\,\Big[{\epsilon_n\cdot \hat{f}_s\cdot\partial_{\epsilon_n}\over \hat{s}_{ns}}\,(\hat{k}_s\cdot\partial_{k_n})^{m-1}-{\epsilon_1\cdot \hat{f}_s\cdot\partial_{\epsilon_1}\over \hat{s}_{s1}}\,(\hat{k}_s\cdot\partial_{k_1})^{m-1}\Big]\,{\cal A}_{\rm YM}(1,\cdots,n)\nn
& &+{\cal R}^{(m-1)}\,.~~\label{high-YM-step1}
\eea
Then, by using \eref{J-YMphi-4p} and \eref{expan-J-np-YM}, one can also find the generalization of \eref{2split-YM-sub} as
\bea
{\cal A}^{(m)}_{\rm YM}(1,\cdots,n,s)\,&\xrightarrow[B=\{3,\cdots,n-1\}]{k_s\cdot k_b=0\,,~\eref{condi-polar}}&\,\Big({\epsilon_s\cdot k_n\over \hat{s}_{ns}}-{\epsilon_s\cdot k_1\over \hat{s}_{s1}}\Big)\,\times\,{\tau^{m-1}\over m!}\,(\hat{k}_s\cdot\partial_{k_1})^m\,{\cal A}_{\rm YM}(1,\cdots,n)\,,\nn
{\cal A}^{(m)}_{\rm YM}(1,\cdots,n,s)\,&\xrightarrow[B=\{2,\cdots,n-2\}]{k_s\cdot k_b=0\,,~\eref{condi-polar}}&\,\Big({\epsilon_s\cdot k_n\over \hat{s}_{ns}}-{\epsilon_s\cdot k_1\over \hat{s}_{s1}}\Big)\,\times\,{\tau^{m-1}\over m!}\,(\hat{k}_s\cdot\partial_{k_n})^m\,{\cal A}_{\rm YM}(1,\cdots,n)\,.~~\label{2split-YM-high}
\eea
Comparing \eref{2split-YM-high} with \eref{high-YM-step1}, we get the following analogue of \eref{R0-constrain-YM}
\bea
{\cal R}^{(m-1)}\,\xrightarrow[B=\{3,\cdots,n-1\}]{k_s\cdot k_b=0\,,~\eref{condi-polar}}&&\,{\tau^{m-1}\over m!}\,{\epsilon_s\cdot k_n\over \hat{s}_{ns}}\,\Big[(\hat{k}_s\cdot\partial_{k_1})^m-(\hat{k}_s\cdot\partial_{k_n})^m\Big]\,{\cal A}_{\rm YM}(1,\cdots,n)\nn
=&&\,{\tau^{m-1}\over m!}\,{\epsilon_s\cdot k_n\over \hat{s}_{ns}}\,\hat{k}_s\cdot(\partial_{k_1}-\partial_{k_n})\,\Big[\sum_{q=0}^{m-1}\,(\hat{k}_s\cdot\partial_{k_1})^q
\,(\hat{k}_s\cdot\partial_{k_n})^{m-1-q}\Big]\,{\cal A}_{\rm YM}(1,\cdots,n)\,,\nn
{\cal R}^{(m-1)}\,\xrightarrow[B=\{2,\cdots,n-2\}]{k_s\cdot k_b=0\,,~\eref{condi-polar}}&&\,{\tau^{m-1}\over m!}\,{\epsilon_s\cdot k_1\over \hat{s}_{s1}}\,\Big[(\hat{k}_s\cdot\partial_{k_1})^m-(\hat{k}_s\cdot\partial_{k_n})^m\Big]\,{\cal A}_{\rm YM}(1,\cdots,n)\nn
=&&\,{\tau^{m-1}\over m!}\,{\epsilon_s\cdot k_1\over \hat{s}_{s1}}\,\hat{k}_s\cdot(\partial_{k_1}-\partial_{k_n})\,\Big[\sum_{q=0}^{m-1}\,(\hat{k}_s\cdot\partial_{k_1})^q
\,(\hat{k}_s\cdot\partial_{k_n})^{m-1-q}\Big]\,{\cal A}_{\rm YM}(1,\cdots,n)\,.~~\label{Rm-constrain-YM}
\eea
Using the previous results in \eref{solu-c} and \eref{es-dot-kn}, we observe
\bea
{\cal R}^{(m-1)}\,& &\xrightarrow[B=\{3,\cdots,n-1\}~{\rm or}~B=\{2,\cdots,n-2\}]{k_s\cdot k_b=0\,,~\eref{condi-polar}}\nn
& &~~
{\tau^{m-1}\over 2\,m!}\,\epsilon_s\cdot(\partial_{k_1}-\partial_{k_n})\,\Big[\sum_{q=0}^{m-1}\,(\hat{k}_s\cdot\partial_{k_1})^q
\,(\hat{k}_s\cdot\partial_{k_n})^{m-1-q}\Big]\,{\cal A}_{\rm YM}(1,\cdots,n)\,,
\eea
which naturally yields the conjecture
\bea
{\cal R}^{(m-1)}\,=\,
{\tau^{m-1}\over 2\,m!}\,\epsilon_s\cdot(\partial_{k_1}-\partial_{k_n})\,\Big[\sum_{q=0}^{m-1}\,(\hat{k}_s\cdot\partial_{k_1})^q
\,(\hat{k}_s\cdot\partial_{k_n})^{m-1-q}\Big]\,{\cal A}_{\rm YM}(1,\cdots,n)\,.~~\label{Rm-result-YM}
\eea
Similar as in the ${\rm Tr}(\phi^3)$ case, the effect of above manipulation is the same as in the sub-leading order, thus does not detect further information beyond the sub-leading order. This is one of reasons why the resulting soft theorem is restricted in the subspace.

In summary, the soft theorem of tree-level YM amplitudes at the arbitrary $m^{\rm th}$ order, in the subspace defined  by $k_s\cdot k_b=0$ and \eref{condi-polar} with $b\in\{3,\cdots,n-2\}$, is given as
\bea
{\cal A}^{(m)}_{\rm YM}(1,\cdots,n,s)\,=\,\tau^{m-1}\,S^{(m)}_{\rm YM}\,{\cal A}_{\rm YM}(1,\cdots,n)\,,~~\label{softtheo-YM-high}
\eea
where
\bea
S^{(m)}_{\rm YM}&=&{1\over m!}\,\Big[{\epsilon_s\cdot k_n\over \hat{s}_{ns}}\,(\hat{k}_s\cdot\partial_{k_n})^m-{\epsilon_s\cdot k_1\over \hat{s}_{s1}}\,(\hat{k}_s\cdot\partial_{k_1})^m\Big]\nn
& &-{1\over(m-1)!}\,\Big[{\epsilon_n\cdot \hat{f}_s\cdot\partial_{\epsilon_n}\over \hat{s}_{ns}}\,(\hat{k}_s\cdot\partial_{k_n})^{m-1}-{\epsilon_1\cdot \hat{f}_s\cdot\partial_{\epsilon_1}\over \hat{s}_{s1}}\,(\hat{k}_s\cdot\partial_{k_1})^{m-1}\Big]\nn
& &+{1\over 2\,m!}\,\epsilon_s\cdot(\partial_{k_1}-\partial_{k_n})\,\Big[\sum_{q=0}^{m-1}\,(\hat{k}_s\cdot\partial_{k_1})^q
\,(\hat{k}_s\cdot\partial_{k_n})^{m-1-q}\Big]\,.~~\label{softfac-YM-high}
\eea
%

\subsection{Consistency check}
\label{subsec-check-YM}

In this subsection, we discuss evidences for the conjectured general soft theorem in \eref{softtheo-YM-high} and \eref{softfac-YM-high} in subspace. We will show that such soft theorem is consistent with the gauge invariance and momentum conservation. Furthermore, under the action of transmutation operator which converts YM amplitudes to ${\rm Tr}(\phi^3)$ ones, the soft factor of YM amplitudes in \eref{softfac-YM-high} is turned to the soft factor of ${\rm Tr}(\phi^3)$ amplitudes.

We first check the gauge invariance, which states that the amplitude ${\cal A}_{\rm YM}(1,\cdots,n,s)$ vanishes under the replacement $\epsilon_s\to \hat{k}_s$. The verification is straightforward. After replacing $\epsilon_s$ by $\hat{k}_s$, the first line of \eref{softfac-YM-high} cancels the third one, while the second line vanishes due to the definition of strength tensor $\hat{f}_s^{\mu\nu}$. Thus, we have $S^{(m)}_{\rm YM}\big|_{\epsilon_s\to \hat{k}_s}=0$,
therefore ${\cal A}^{(m)}_{\rm YM}(1,\cdots,n,s)$ in \eref{softtheo-YM-high} vanishes for arbitrary $m$.

Then we verify the consistency with momentum conservation. According to the argument the same as in section \ref{subsec-check-phi3}, soft operators for the YM case should satisfy the following commutation relation
\bea
[S^{(m)}_{\rm YM}\,,\,P_n]\,{\cal A}_{\rm YM}(1,\cdots,n)=\big(\hat{k}_s\cdot V\big)\,S^{(m-1)}_{\rm YM}\,{\cal A}_{\rm YM}(1,\cdots,n)\,,~~\label{commu-momen-conser-YM}
\eea
which is analogous to \eref{commu-momen-conser}, where $P_n$ and $V$ are defined in \eref{defin-P}. To verify \eref{commu-momen-conser-YM}, we notice that the observations \eref{veri-mom-obser0}, \eref{veri-mom-obser1} and \eref{veri-mom-obser2} in section \ref{subsec-check-phi3} are also correct for the YM case, with ${\cal A}_{{\rm Tr}(\phi^3)}$ being replaced by ${\cal A}_{\rm YM}$. Using these observations, as well as the definition of $S^{(m)}_{\rm YM}$ in \eref{softfac-YM-high}, we immediately obtain
\bea
& &[S^{(m)}_{\rm YM}\,,\,P_n]\,{\cal A}_{\rm YM}(1,\cdots,n)\nn
&=&{\hat{k}_s\cdot V\over (m-1)!}\,\Big[{\epsilon_s\cdot k_n\over \hat{s}_{ns}}\,(\hat{k}_s\cdot\partial_{k_n})^{m-1}-{\epsilon_s\cdot k_1\over \hat{s}_{s1}}\,(\hat{k}_s\cdot\partial_{k_1})^{m-1}\Big]\,{\cal A}_{\rm YM}(1,\cdots,n)\nn
& &-{\hat{k}_s\cdot V\over(m-2)!}\,\Big[{\epsilon_n\cdot \hat{f}_s\cdot\partial_{\epsilon_n}\over \hat{s}_{ns}}\,(\hat{k}_s\cdot\partial_{k_n})^{m-2}-{\epsilon_1\cdot \hat{f}_s\cdot\partial_{\epsilon_1}\over \hat{s}_{s1}}\,(\hat{k}_s\cdot\partial_{k_1})^{m-2}\Big]\,{\cal A}_{\rm YM}(1,\cdots,n)\nn
& &+{\hat{k}_s\cdot V\over 2\,(m-1)!}\,\epsilon_s\cdot(\partial_{k_1}-\partial_{k_n})\,\Big[\sum_{q=0}^{m-2}\,(\hat{k}_s\cdot\partial_{k_1})^q
\,(\hat{k}_s\cdot\partial_{k_n})^{m-2-q}\Big]\,{\cal A}_{\rm YM}(1,\cdots,n)\nn
&=&\big(\hat{k}_s\cdot V\big)\,S^{(m-1)}_{\rm YM}\,{\cal A}_{\rm YM}(1,\cdots,n)\,,
\eea
where we have used the computation in \eref{operator-D},
\bea
& &\sum_{q=0}^{m-1}\,\Big(q\,(\hat{k}_s\cdot\partial_{k_1})^{q-1}\,(\hat{k}_s\cdot\partial_{k_n})^{m-1-q}+(m-1-q)\,
(\hat{k}_s\cdot\partial_{k_1})^q\,(\hat{k}_s\cdot\partial_{k_n})^{m-2-q}
\Big)\nn
&=&\sum_{q'=0}^{m-2}\,(q'+1)\,(\hat{k}_s\cdot\partial_{k_1})^{q'}\,(\hat{k}_s\cdot\partial_{k_n})^{m-2-q'}+\sum_{q=0}^{m-2}\,(m-1-q)\,
(\hat{k}_s\cdot\partial_{k_1})^q\,(\hat{k}_s\cdot\partial_{k_n})^{m-2-q}\nn
&=&\sum_{q=0}^{m-2}\,m\,(\hat{k}_s\cdot\partial_{k_1})^{q}\,(\hat{k}_s\cdot\partial_{k_n})^{m-2-q}.~~\label{ana-compuD}
\eea
As mentioned below \eref{operator-D}, $(\hat{k}_s\cdot\partial_{k_1})^{q-1}$ in the first line of \eref{ana-compuD}
should be understood as $0$ when $q=0$, and the analogous understanding holds for $(\hat{k}_s\cdot\partial_{k_n})^{m-2-q}$ when $q=m-1$.

Finally, we verify that the transmutation operator proposed in \cite{Cheung:2017ems}, which turns YM amplitudes to ${\rm Tr}(\phi^3)$ ones, does convert the conjectured soft operator $S^{(m)}_{\rm YM}$ to the corresponding operator $S^{(m)}_{{\rm Tr}(\phi^3)}$. Such transmutation operator, satisfying
\bea
{\cal T}(1,\cdots,n,s)\,{\cal A}_{\rm YM}(1,\cdots,n,s)\,=\,{\cal A}_{{\rm Tr}(\phi^3)}(1,\cdots,n,s)\,,~~\label{trans-YM}
\eea
is defined as
\bea
{\cal T}(1,\cdots,n,s)\,\equiv\,{\cal I}_{ns1}\,{\cal T}(1,\cdots,n)\,,~~~~~~{\cal T}(1,\cdots,n)\,\equiv\,\partial_{\epsilon_1\cdot\epsilon_n}\,\prod_{i=2}^{n-1}\,{\cal I}_{(i-1)in}\,,
\eea
where the insertion operators are defined as
\bea
{\cal I}_{ns1}\,\equiv\,\partial_{\epsilon_s\cdot k_n}-\partial_{\epsilon_s\cdot k_1}\,,~~~~~~{\cal I}_{(i-1)in}\,\equiv\,
\partial_{\epsilon_i\cdot k_{i-1}}-\partial_{\epsilon_i\cdot k_n}\,.~~\label{defin-inser}
\eea
The operator ${\cal T}(1,\cdots,n)$ is the transmutation operator for $n$-point amplitudes, which transmutes the $n$-point YM amplitude via the same pattern as in \eref{trans-YM}, namely,
\bea
{\cal T}(1,\cdots,n)\,{\cal A}_{\rm YM}(1,\cdots,n)={\cal A}_{{\rm Tr}(\phi^3)}(1,\cdots,n)\,. ~~\label{trans-YM-phi}
\eea
Now we act the operator ${\cal T}(1,\cdots,n,s)$ on the soft behavior of ${\cal A}_{\rm YM}(1,\cdots,n,s)$ in \eref{softtheo-YM-high}. Since the operator ${\cal T}(1,\cdots,n,s)$ chosen above is independent of $k_s$ which carries the soft parameter $\tau$, we expect that this operator turns ${\cal A}^{(m)}_{\rm YM}(1,\cdots,n,s)$ to ${\cal A}^{(m)}_{{\rm Tr}(\phi^3)}(1,\cdots,n,s)$ with the same $m$, thus let us check this. By definition, the operator ${\cal T}(1,\cdots,n)$ is commutable with
$S^{(m)}_{\rm YM}$, thus we can substitute \eref{trans-YM-phi} into \eref{softtheo-YM-high}, to obtain
\bea
& &{\cal T}(1,\cdots,n,s)\,{\cal A}^{(m)}_{\rm YM}(1,\cdots,n,s)\nn
&=&{\cal I}_{ns1}\,S^{(m)}_{\rm YM}\,{\cal A}_{{\rm Tr}(\phi^3)}(1,\cdots,n)\nn
&=&{1\over m!}\,\Big[{1\over \hat{s}_{ns}}\,(\hat{k}_s\cdot\partial_{k_n})^m+{1\over \hat{s}_{s1}}\,(\hat{k}_s\cdot\partial_{k_1})^m\Big]\,{\cal A}_{{\rm Tr}(\phi^3)}(1,\cdots,n)\nn
& &+{1\over 2\,m!}\,\Big[\sum_{q=0}^{m-1}\,(\hat{k}_s\cdot\partial_{k_1})^q
\,(\hat{k}_s\cdot\partial_{k_n})^{m-1-q}\Big]\,{\cal I}_{ns1}\,\epsilon_s\cdot(\partial_{k_1}-\partial_{k_n})\,{\cal A}_{{\rm Tr}(\phi^3)}(1,\cdots,n)\,,
~~\label{trans-YM-step}
\eea
where the second equality uses the definitions of ${\cal I}_{ns1}$ and $S^{(m)}_{\rm YM}$ in \eref{defin-inser} and \eref{softfac-YM-high}, respectively. Substituting the result in \eref{deri-phi3} into the last line of \eref{trans-YM-step}, we arrive at
\bea
& &{\cal T}(1,\cdots,n,s)\,{\cal A}^{(m)}_{\rm YM}(1,\cdots,n,s)\nn
&=&{1\over m!}\,\Big[{1\over \hat{s}_{ns}}\,(\hat{k}_s\cdot\partial_{k_n})^m+{1\over \hat{s}_{s1}}\,(\hat{k}_s\cdot\partial_{k_1})^m\Big]\,{\cal A}_{{\rm Tr}(\phi^3)}(1,\cdots,n)\nn
& &-{1\over m!}\,\Big[\sum_{q=0}^{m-1}\,(\hat{k}_s\cdot\partial_{k_1})^q
\,(\hat{k}_s\cdot\partial_{k_n})^{m-1-q}\Big]\,\Big(\sum_{i=2}^{n-2}\,\partial_{s_{1\cdots i}}\Big)\,{\cal A}_{{\rm Tr}(\phi^3)}(1,\cdots,n)\nn
&=&{\cal A}^{(m)}_{{\rm Tr}(\phi^3)}(1,\cdots,n,s)\,,~~\label{trans-YM-result}
\eea
where the last equality uses the soft theorem of ${\rm Tr}(\phi^3)$ amplitudes given in \eref{softtheo-phi3-high} and \eref{softfac-phi3-high}.
The above relation \eref{trans-YM-result} indicates that the transmutation operator turns soft theorems of YM amplitudes in the subspace to soft theorems of ${\rm Tr}(\phi^3)$ amplitudes in the subspace, where the subspace for ${\rm Tr}(\phi^3)$ is embedded in the subspace for YM, and one can reduce the subspace for YM to that for ${\rm Tr}(\phi^3)$ by stripping off polarizations. The transmutation relation \eref{trans-YM-result} supports the conjectured soft theorem \eref{softtheo-YM-high} of YM amplitudes at the $m^{\rm th}$ order, with the soft factor given in \eref{softfac-YM-high}.

\section{Double-soft theorems of NLSM amplitudes}
\label{sec-NLSM}

In this section, we study the double-soft theorems of tree-level NLSM amplitudes at leading and sub-leading orders, where two external pions are taken to be soft. The soft pions are labeled as $s_1$ and $s_2$, with their corresponding momenta parameterized as $k_{s_1}\to\tau\hat{k}_{s_1}$, $k_{s_2}\to\tau\hat{k}_{s_2}$, in the limit $\tau\to 0$. We restrict our analysis to the special case where $s_1$ and $s_2$ are consecutive in the color ordering, i.e., $s_2=s_1+1$.

For the leading double-soft theorem of NLSM amplitudes, a new situation arises. As will be explained in subsection \ref{subsec-leading-NLSM}, this leading soft theorem cannot be derived by considering only Feynman diagrams with divergent propagators. However, the method utilizing the 2-split in \eref{2split-NLSM-special} with $A=\{s_1,s_2\}$, still yields the well defined complete leading soft theorem. In subsection \ref{subsec-higher-NLSM}, we demonstrate that the method developed in sections \ref{subsec-subleading-phi3} and \ref{subsec-subleading-YM} can also determine the sub-leading double-soft theorem for NLSM. Subsequently, in subsection \ref{subsec-veri-NLSM}, we employ the more general $2$-split from \eref{2split-NLSM-general} to verify these results, and identify the sub-leading soft behavior of the pure NLSM current in this $2$-split. An interesting observation is, the replacement $\epsilon_s\to k_{s_1}-k_{s_2}$, $k_s\to k_{s_1}+k_{s_2}$ not only connects the soft behaviors of YM and NLSM amplitudes, but also converts gauge invariance in YM to Adler zero in NLSM. Finally, we note that while the method in subsection \ref{subsec-higher-NLSM} can be extended to derive higher order soft theorems in kinematic subspaces, the derivation becomes considerably more involved, and a universal representation of the higher order soft factor remains unclear.

\subsection{Leading order}
\label{subsec-leading-NLSM}

%
\begin{figure}
  \centering
   \includegraphics[width=4.3cm]{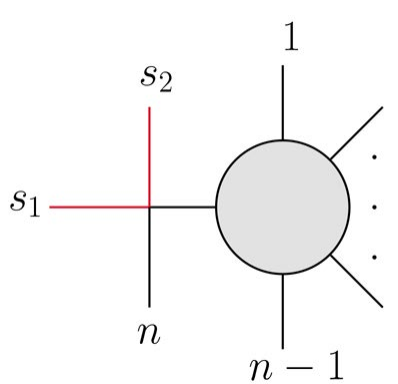}
   ~~~~~~~~~~~~~~~~~~~~
   \includegraphics[width=4.3cm]{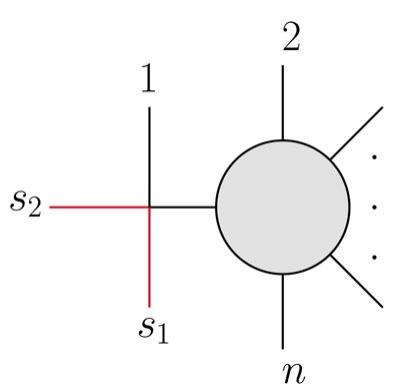} \\
  \caption{Diagrams with divergent propagators in the double-soft limit, where the soft particles represented in red are attached to the quadrivalent vertex in each diagram.}\label{pole-part-NLSM}
\end{figure}

For ${\rm Tr}(\phi^3)$ and YM amplitudes studied in section \ref{sec-phi3} and section \ref{sec-YM}, their leading single-soft theorems can be derived by utilizing either usual factorizations or $2$-splits. However, for double-soft behavior of NLSM amplitudes under consideration in this subsection, the first approach is not effective. The reason is as follows. Let us consider diagrams in Fig.\ref{pole-part-NLSM}, i.e., diagrams contain propagators $1/ s_{ns_1s_2}$ or $1/s_{s_1s_21}$ which are divergent in the double soft limit. For the left diagram, the factorization at the pole $s_{ns_1s_2}=0$ gives rise to the $4$-point sub-amplitude ${\cal A}_{\rm NLSM}(n,s_1,s_2,I)$ at the leading order, as well as another $n-2$-point one. The $4$-point amplitude ${\cal A}_{\rm NLSM}(n,s_1,s_2,I)$ can be evaluated via Feynman rules, or bootstrapped via mass dimension and cyclic symmetry of four external pions, resulted in
\bea
{\cal A}_{\rm NLSM}(n,s_1,s_2,I)={\tau\,\hat{k}_{s_2}\cdot k_n\over2}\,.~~\label{4p-NLSM-1}
\eea
The soft parameter $\tau$ carried by ${\cal A}_{\rm NLSM}(n,s_1,s_2,I)$ cancels $\tau^{-1}$ from the propagator $1/s_{ns_1s_2}$. It means the leading term contributed by the left diagram in Fig.\ref{pole-part-NLSM} is at the $\tau^0$ order. The similar argument and conclusion hold for the right diagram in Fig.\ref{pole-part-NLSM}. Since diagrams without divergent propagators can also contribute to the $\tau^0$ order, we know that the leading soft behavior of NLSM amplitudes does not solely arise from diagrams in Fig.\ref{pole-part-NLSM}. Therefore, computing on-shell $4$-point amplitudes in factorizations, and combining them with divergent propagators, are not sufficient to determine the leading soft theorem.

However, the method based on $2$-splits is still valid in the NLSM case. Choosing $A=\{s_1,s_2\}$, $B=\{3,\cdots,n-1\}$ or $B=\{2,\cdots,n-2\}$,
the $n+2$-point NLSM amplitude satisfies $2$-splits as in \eref{2split-NLSM-special},
\bea
{\cal A}_{\rm NLSM}(1,\cdots,n,s_1,s_2)\,&\xrightarrow[B=\{3,\cdots,n-1\}]{k_a\cdot k_b=0}&\,{\cal J}_{{\rm NLSM}\oplus{\rm Tr}(\phi^3)}(n_\phi,s_{1p},s_{2p},1_\phi,2^*_\phi)\,\times\,{\cal J}_{\rm NLSM}(1^*,\cdots,n)\,,\nn
{\cal A}_{\rm NLSM}(1,\cdots,n,s_1,s_2)\,&\xrightarrow[B=\{2,\cdots,n-2\}]{k_a\cdot k_b=0}&\,{\cal J}_{{\rm NLSM}\oplus{\rm Tr}(\phi^3)}(n-1^*_\phi,n_\phi,s_{1p},s_{2p},1_\phi)\,\times\,{\cal J}_{\rm NLSM}(1,\cdots,n^*)\,,~~\label{2split-NLSM-soft}
\eea
for $a\in A$, $b\in B$. In the above, each NLSM$\oplus{\rm Tr}(\phi^3)$ current ${\cal J}_{{\rm NLSM}\oplus{\rm Tr}(\phi^3)}$ carries two external pions $s_{1p}$, $s_{2p}$, as well as three external scalars of ${\rm Tr}(\phi^3)$ theory labeled by subscripts $\phi$. Since the leading contribution of ${\cal J}_{\rm NLSM}(1,\cdots,n^*)$ is ${\cal A}_{\rm NLSM}(1,\cdots,n)$, and the double-soft limit is compatible with $k_a\cdot k_b=0$, we conclude that the leading contribution of NLSM$\oplus{\rm Tr}(\phi^3)$
currents automatically incorporates the desired soft factor.

\begin{figure}
  \centering
   \includegraphics[width=12cm]{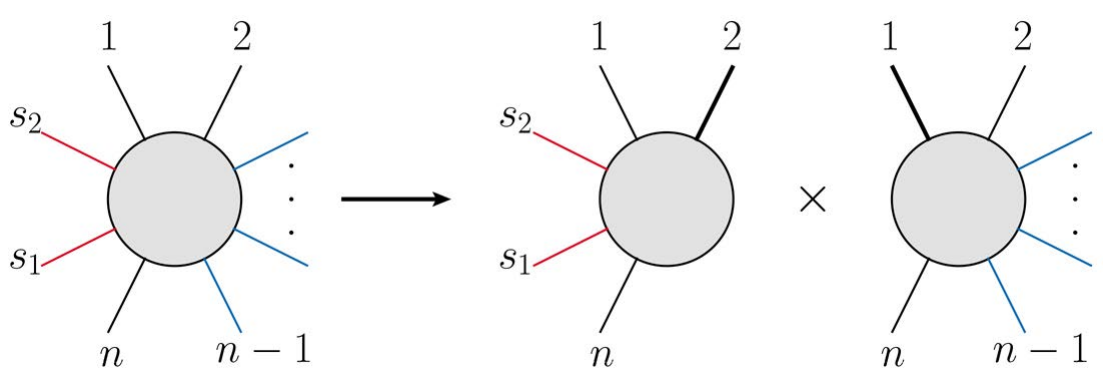}\\
   ~~~~~~~~\\
   ~~~~~~~~\\
   \includegraphics[width=12cm]{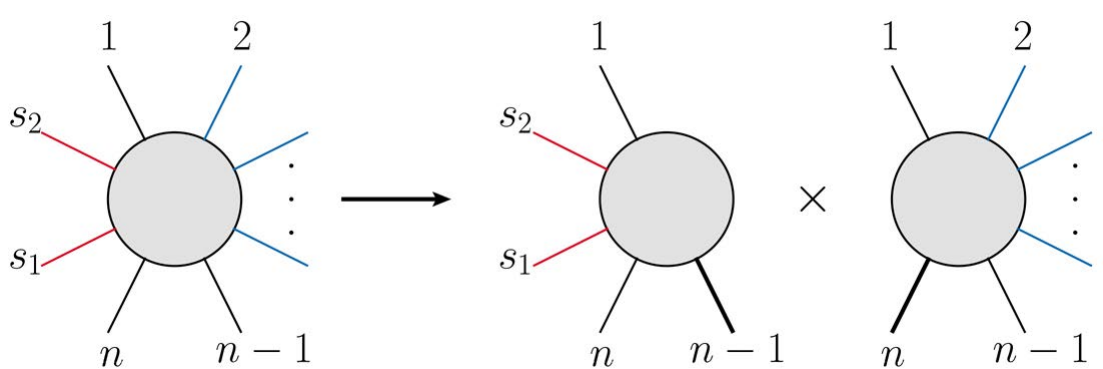} \\
  \caption{Two specific $2$-splits for NLSM amplitudes, where the set $A$ includes two external legs $s_1$ and $s_2$ represented in red, while legs in $B$ are represented in blue. The bold lines again represent off-shell particles.}\label{split-part-NLSM}
\end{figure}

Let us figure out the leading soft theorem along the line discussed above.
We can use momentum conservation to remove $k_2^*$ in ${\cal J}_{{\rm NLSM}\oplus{\rm Tr}(\phi^3)}(n_\phi,s_{1p},s_{2p},1_\phi,2^*_\phi)$ and $k_{n-1}^*$ in ${\cal J}_{{\rm NLSM}\oplus{\rm Tr}(\phi^3)}(n-1^*_\phi,n_\phi,s_{1p},s_{2p},1_\phi)$, then find
\bea
& &{\cal J}_{{\rm NLSM}\oplus{\rm Tr}(\phi^3)}(n_\phi,s_{1p},s_{2p},1_\phi,2^*_\phi)\nn
&=&{\cal J}_{{\rm NLSM}\oplus{\rm Tr}(\phi^3)}(n-1^*_\phi,n_\phi,s_{1p},s_{2p},1_\phi)\nn
&=&{(k_{s_1}-k_{s_2})\cdot k_n+\tau\,k_{s_1}\cdot k_{s_2}\over 2\,s_{ns_1s_2}}+{(k_{s_2}-k_{s_1})\cdot k_1+\tau\,k_{s_1}\cdot k_{s_2}\over 2\,s_{s_1s_21}}\,.~~\label{J-5p-NLSM+phi}
\eea
Extracting leading terms in \eref{J-5p-NLSM+phi}, we obtain the leading soft theorem of NLSM amplitudes
\bea
{\cal A}^{(0)}_{\rm NLSM}(1,\cdots,n,s_1,s_2)=\Big[{(\hat{k}_{s_1}-\hat{k}_{s_2})\cdot k_n\over 2\,\hat{s}_{ns_1s_2}}+{(\hat{k}_{s_2}-\hat{k}_{s_1})\cdot k_1\over 2\,\hat{s}_{s_1s_21}}\Big]\,{\cal A}_{\rm NLSM}(1,\cdots,n)\,,~~\label{softtheo-NLSM-leading}
\eea
the same as the standard one in \cite{Cachazo:2015ksa,Du:2015esa}.
Here we still parameterize the divergent propagators as $1/s_{ns_1s_2}\to1/\tau\hat{s}_{ns_1s_2}$, $1/s_{s_1s_21}\to1/\tau\hat{s}_{s_1s_21}$, similar as in previous sections. However, by definition, we have
\bea
{1\over\hat{s}_{ns_1s_2}}&=&{1\over2\,(\hat{k}_{s_1}+\hat{k}_{s_2})\cdot k_n+2\,\tau\,\hat{k}_{s_1}\cdot\hat{k}_{s_2}}\,,\nn
{1\over\hat{s}_{s_1s_21}}&=&{1\over2\,(\hat{k}_{s_1}+\hat{k}_{s_2})\cdot k_1+2\,\tau\,\hat{k}_{s_1}\cdot\hat{k}_{s_2}}\,,~~\label{propagat-diver-NLSM}
\eea
thus they contain terms at higher orders of $\tau$. It is more convenient and compact to describe soft behaviors at leading and sub-leading orders by using these $\hat{s}_{ns_1s_2}$ and $\hat{s}_{s_1s_21}$ as in \eref{softtheo-NLSM-leading}. However, one should keep in mind the formula in \eref{softtheo-NLSM-leading} also contains contribution to higher orders.

Interestingly, the leading double-soft theorem of NLSM amplitudes in \eref{softtheo-NLSM-leading}, can be convert to the leading single-soft theorem of YM amplitudes in \eref{softtheo-YM-leading} up to an overall $1/2$, by replacing
\bea
\hat{k}_{s_1}-\hat{k}_{s_2}\to \epsilon_s\,,~~~~~~~~\hat{k}_{s_1}+\hat{k}_{s_2}\to \hat{k}_s\,,~~\label{replace}
\eea
and neglecting terms in \eref{propagat-diver-NLSM} which do not contribute to the leading order.
This phenomenon may has deep connection to the discovery in \cite{Arkani-Hamed:2023jry,Arkani-Hamed:2024nhp}: the stingy description of $n$-point YM amplitudes can be obtained from the stringy formula of $2n$-point ${\rm Tr}(\phi^3)$ amplitudes via the scaffolding technic based on the above replacement, and the NLSM amplitudes is contained in the stringy formula of ${\rm Tr}(\phi^3)$.

\subsection{Sub-leading order}
\label{subsec-higher-NLSM}

As can be seen in sections \ref{subsec-subleading-phi3} and \ref{subsec-subleading-YM}, separations of amplitudes in \eref{decom-phi3} and \eref{decom-YM} are useful when deriving sub-leading soft theorems of ${\rm Tr}(\phi^3)$ and YM amplitudes. Thus, to apply this approach to the NLSM case, we can begin with the seeking for the analogous separation. In each of \eref{decom-phi3} and \eref{decom-YM}, the first part corresponds to diagrams in Fig.\ref{pole-part}, while the remaining part corresponds to other diagrams. However, for the NLSM case, the separation based on such diagrammatic understanding is not the most efficient choice. As studied in the previous subsection, the leading soft behavior receives contributions from not only diagrams in Fig.\ref{pole-part-NLSM}. Thus, the leading soft theorem \eref{softtheo-NLSM-leading} tells us, other diagrams also provide ${\cal J}_{\rm NLSM}(1,\cdots,n^*)/s_{ns_1s_2}$ or ${\cal J}_{\rm NLSM}(1^*,\cdots,n)/s_{s_1s_21}$. Hence, we can consider the the following separation formula
\bea
{\cal A}_{\rm NLSM}(1,\cdots,n,s_1,s_2)&=&\Big[\W{\cal J}_{\rm NLSM}(s_1,s_2,1,I_1)\,{1\over s_{s_1s_21}}\,{\cal J}_{\rm NLSM}(1^*,\cdots,n)\nn
& &+\W{\cal J}_{\rm NLSM}(n,s_1,s_2,I_2)\,{1\over s_{ns_1s_2}}\,{\cal J}_{\rm NLSM}(1,\cdots,n^*)\Big]\,+\,{\cal R}\,,~~\label{decom-NLSM}
\eea
where the notation $\W{\cal J}_{\rm NLSM}$ is introduced to emphasize that we do not require these currents to correspond to Fig.\ref{pole-part-NLSM}. There are various degrees of freedom for choosing $\W{\cal J}_{\rm NLSM}(s_1,s_2,1,I_1)$ and $\W{\cal J}_{\rm NLSM}(n,s_1,s_2,I_2)$. Firstly, Feynman rules for vertices sensitively depend on the parameterization, as studied in \cite{Kampf:2013vha}. Secondly, if ${\cal R}$ includes a term proportional to ${\cal J}_{\rm NLSM}(1^*,\cdots,n)$ (or ${\cal J}_{\rm NLSM}(1,\cdots,n^*)$), one can always trivially turn it to $(s_{s_1s_21}/s_{s_1s_21}){\cal J}_{\rm NLSM}(1^*,\cdots,n)$ (or $(s_{ns_1s_2}/s_{ns_1s_2}){\cal J}_{\rm NLSM}(1,\cdots,n^*)$), and move it to the first part by absorbing the numerator $s_{s_1s_21}$ (or $s_{ns_1s_2}$) into $\W{\cal J}_{\rm NLSM}(s_1,s_2,1,I_1)$ (or $\W{\cal J}_{\rm NLSM}(n,s_1,s_2,I_2)$). Nevertheless, we can require the separation \eref{decom-NLSM} to incorporate another significant feature of separations \eref{decom-phi3} and \eref{decom-YM}, i.e., ${\cal R}$ in \eref{decom-NLSM} does not contribute to the leading soft behavior. Thus, $\W{\cal J}_{\rm NLSM}(s_1,s_2,1,I_1)$ and $\W{\cal J}_{\rm NLSM}(n,s_1,s_2,I_2)$ should satisfy
\bea
\W{\cal J}^{(0)}_{\rm NLSM}(s_1,s_2,1,I_1)&=&{\tau\over2}\,(\hat{k}_{s_2}-\hat{k}_{s_1})\cdot k_1\,,\nn
\W{\cal J}^{(0)}_{\rm NLSM}(n,s_1,s_2,I_2)&=&{\tau\over2}\,(\hat{k}_{s_1}-\hat{k}_{s_2})\cdot k_n\,,
\eea
due to the leading soft theorem \eref{softtheo-NLSM-leading}.
$\W{\cal J}_{\rm NLSM}(s_1,s_2,1,I_1)$ and $\W{\cal J}_{\rm NLSM}(n,s_1,s_2,I_2)$ are also allowed to contain terms at the $\tau^2$ order, namely,
\bea
\W{\cal J}_{\rm NLSM}(s_1,s_2,1,I_1)&=&{\tau\over2}\,(\hat{k}_{s_2}-\hat{k}_{s_1})\cdot k_1+\tau^2\,\a_1\,\hat{k}_{s_1}\cdot \hat{k}_{s_2}\,,\nn
\W{\cal J}_{\rm NLSM}(n,s_1,s_2,I_2)&=&{\tau\over2}\,(\hat{k}_{s_1}-\hat{k}_{s_2})\cdot k_n+\tau^2\,\a_n\,\hat{k}_{s_1}\cdot \hat{k}_{s_2}\,.~~\label{choice-J-4p-boot}
\eea
Any higher order is forbidden, due to the mass dimension.

The $\tau^2$ terms in \eref{choice-J-4p-boot} will contribute to the sub-leading soft behavior. Similar as in \eref{YM-subleading}, we can use the separation \eref{decom-NLSM} to express the sub-leading soft behavior of NLSM amplitude as
\bea
& &{\cal A}^{(1)}_{\rm NLSM}(1,\cdots,n,s_1,s_2)\nn
&=&{1\over\tau}\,\W{\cal J}^{(0)}_{\rm NLSM}(s_1,s_2,1,I_1)\,{1\over \hat{s}_{s_1s_21}}\,{\cal J}^{(1)}_{\rm NLSM}(1^*,\cdots,n)+{1\over\tau}\,\W{\cal J}^{(0)}_{\rm NLSM}(n,s_1,s_2,I_2)\,{1\over \hat{s}_{ns_1s_2}}\,{\cal J}^{(1)}_{\rm NLSM}(1,\cdots,n^*)\nn
& &+{1\over\tau}\,\W{\cal J}^{(1)}_{\rm NLSM}(s_1,s_2,1,I_1)\,{1\over \hat{s}_{s_1s_21}}\,{\cal J}^{(0)}_{\rm NLSM}(1^*,\cdots,n)+{1\over\tau}\,\W{\cal J}^{(1)}_{\rm NLSM}(n,s_1,s_2,I_2)\,{1\over \hat{s}_{ns_1s_2}}\,{\cal J}^{(0)}_{\rm NLSM}(1,\cdots,n^*)\nn
& &+\,{\cal R}^{(0)}\,.~~\label{NLSM-subleading}
\eea
Using the ansatz in \eref{choice-J-4p-boot}, we have
\bea
\W{\cal J}^{(1)}_{\rm NLSM}(s_1,s_2,1,I_1)&=&\tau^2\,\a_1\,\hat{k}_{s_1}\cdot \hat{k}_{s_2}\,,\nn
\W{\cal J}^{(1)}_{\rm NLSM}(n,s_1,s_2,I_2)&=&\tau^2\,\a_n\,\hat{k}_{s_1}\cdot \hat{k}_{s_2}\,.~~\label{J-4p-sub-boot}
\eea
To fix $\a_1$ and $\a_n$ in the above, we can further require that ${\cal R}^{(0)}$ in \eref{NLSM-subleading} does not contain $\hat{k}_{s_1}\cdot \hat{k}_{s_2}/\hat{s}_{s_1s_21}$ or $\hat{k}_{s_1}\cdot \hat{k}_{s_2}/\hat{s}_{ns_1s_2}$. Meanwhile, we observe that the special locus $k_a\cdot k_b=0$ in kinematic space has no influence on terms in \eref{NLSM-subleading} which carry $\W{\cal J}^{(1)}_{\rm NLSM}(s_1,s_2,1,I_1)$ or $\W{\cal J}^{(1)}_{\rm NLSM}(n,s_1,s_2,I_2)$, since
\bea
{\cal J}^{(0)}_{\rm NLSM}(1^*,\cdots,n)={\cal J}^{(0)}_{\rm NLSM}(1,\cdots,n^*)={\cal A}_{\rm NLSM}(,\cdots,n)\,.
\eea
Thus, we can use $2$-splits in \eref{2split-NLSM-soft} and the expression of $5$-point currents in \eref{J-5p-NLSM+phi}, to find $\a_1=\a_n=1/2$,
and
\bea
\W{\cal J}_{\rm NLSM}(s_1,s_2,1,I_1)&=&{\tau\over2}\,\Big((\hat{k}_{s_2}-\hat{k}_{s_1})\cdot k_1+\tau\,\hat{k}_{s_1}\cdot \hat{k}_{s_2}\Big)\,,\nn
\W{\cal J}_{\rm NLSM}(n,s_1,s_2,I_2)&=&{\tau\over2}\,\Big((\hat{k}_{s_1}-\hat{k}_{s_2})\cdot k_n+\tau\,\hat{k}_{s_1}\cdot \hat{k}_{s_2}\Big)\,.~~\label{choice-J-4p-fix}
\eea
With the above choice of $\W{\cal J}_{\rm NLSM}(s_1,s_2,1,I_1)$ and $\W{\cal J}_{\rm NLSM}(n,s_1,s_2,I_2)$, the ${\cal R}^{(0)}$ part in \eref{NLSM-subleading} only contributes terms without $1/\hat{s}_{s_1s_21}$ or $1/\hat{s}_{ns_1s_2}$.

With the separation \eref{decom-NLSM} chosen above, we are ready to completely determine the sub-leading double-soft theorem of NLSM amplitudes. Expanding ${\cal J}_{\rm NLSM}(1^*,\cdots,n)$ and ${\cal J}_{\rm NLSM}(1,\cdots,n^*)$ as
\bea
{\cal J}_{\rm NLSM}(1^*,\cdots,n)&=&{\cal A}_{\rm NLSM}(1,\cdots,n)+\sum_{m=1}^\infty\,{\tau^m\over m!}\,\Big((\hat{k}_{s_1}+\hat{k}_{s_2})\cdot\partial_{k_1}\Big)^m\,{\cal A}_{\rm NLSM}(1,\cdots,n)\,,\nn
{\cal J}_{\rm NLSM}(1,\cdots,n^*)&=&{\cal A}_{\rm NLSM}(1,\cdots,n)+\sum_{m=1}^\infty\,{\tau^m\over m!}\,\Big((\hat{k}_{s_1}+\hat{k}_{s_2})\cdot\partial_{k_n}\Big)^m\,{\cal A}_{\rm NLSM}(1,\cdots,n)\,,~~~\label{expan-NLSMcurrent}
\eea
and admitting the choice \eref{choice-J-4p-fix}, one can simplify \eref{NLSM-subleading} as
\bea
& &{\cal A}^{(1)}_{\rm NLSM}(1,\cdots,n,s_1,s_2)\nn
&=&\tau\,\Big[{(\hat{k}_{s_1}-\hat{k}_{s_2})\cdot k_n\over 2\,\hat{s}_{ns_1s_2}}\,(\hat{k}_{s_1}+\hat{k}_{s_2})\cdot\partial_{k_n}+{(\hat{k}_{s_2}-\hat{k}_{s_1})\cdot k_1\over 2\,\hat{s}_{s_1s_21}}\,(\hat{k}_{s_1}+\hat{k}_{s_2})\cdot\partial_{k_1}\Big]\,{\cal A}_{\rm NLSM}(1,\cdots,n)\nn
& &+\tau\,\,\Big({\hat{k}_{s_1}\cdot \hat{k}_{s_2}\over 2\,\hat{s}_{ns_1s_2}}+{\hat{k}_{s_1}\cdot \hat{k}_{s_2}\over 2\,\hat{s}_{s_1s_21}}\Big)\,{\cal A}_{\rm NLSM}(1,\cdots,n)\,+\,{\cal R}^{(0)}\,.~~\label{NLSM-subleading-step1}
\eea

Before proceeding, it is worth to compare the derivation for the above \eref{NLSM-subleading-step1} with the derivation for the ansatz \eref{sub-YM-step1}
in the YM case. The first one utilizes solely $2$-splits, while the second one exploits both $2$-splits and $3$-point currents in \eref{J-3p-YM}.
The reason is as follows. In the YM case, the resulting $4$-point currents in $2$-splits \eref{2split-YM-soft} are YM$\oplus{\rm Tr}(\phi^3)$ ones, each of them carries only one external gluon $s_g$. However, currents ${\cal J}_{\rm YM}(s,1,I_1)$ and ${\cal J}_{\rm YM}(n,s,I_2)$ in the separation \eref{decom-YM} are pure YM ones, each of them carries two on-shell external gluons. Thus, $4$-point currents in\eref{2split-YM-soft} can not capture all information of $3$-point currents in \eref{decom-YM}. The situation is changed in the NLSM case, according to the character that the $m$-point interaction $p-p-p-p-\cdots$ for pure pions is equivalent to the $m$-point interaction $\phi-\phi-p-p-\cdots$ for $m-2$ pions and $2$ scalars of ${\rm Tr}(\phi^3)$ \cite{Low:2017mlh,Mizera:2018jbh}. The above property implies that the NLSM currents $\W{\cal J}_{\rm NLSM}(s_1,s_2,1,I_1)$ and $\W{\cal J}_{\rm NLSM}(n,s_1,s_2,I_2)$ in \eref{decom-NLSM} are equivalent to NLSM$\oplus{\rm Tr}(\phi^3)$ ones $\W{\cal J}_{{\rm NLSM}\oplus{\rm Tr}(\phi^3)}(s_{1p},s_{2p},1_\phi,I_{1\phi})$ and $\W{\cal J}_{{\rm NLSM}\oplus{\rm Tr}(\phi^3)}(n_\phi,s_{1p},s_{2p},I_{2\phi})$, thus, their information are completely captured by $5$-point NLSM$\oplus{\rm Tr}(\phi^3)$ currents in $2$-splits in \eref{2split-NLSM-soft}.

Now we fix ${\cal R}^{(0)}$ in the ansatz \eref{NLSM-subleading-step1} by using $2$-splits. By substituting $5$-point currents \eref{J-5p-NLSM+phi} and expansions \eref{expan-NLSMcurrent} into \eref{2split-NLSM-soft}, we get
\bea
& &{\cal A}^{(1)}_{\rm NLSM}(1,\cdots,n,s_1,s_2)\,\xrightarrow[B=\{3,\cdots,n-1\}]{k_a\cdot k_b=0}\nn
& &~~~~~~~~~~~~~~~~~~~~~~~~\tau\,\Big[{(\hat{k}_{s_1}-\hat{k}_{s_2})\cdot k_n\over 2\,\hat{s}_{ns_1s_2}}+{(\hat{k}_{s_2}-\hat{k}_{s_1})\cdot k_1\over 2\,\hat{s}_{s_1s_21}}\Big]\,(\hat{k}_{s_1}+\hat{k}_{s_2})\cdot\partial_{k_1}\,{\cal A}_{\rm NLSM}(1,\cdots,n)\nn
& &~~~~~~~~~~~~~~~~~~~~~~~~+\tau\,\Big({\hat{k}_{s_1}\cdot \hat{k}_{s_2}\over 2\,\hat{s}_{ns_1s_2}}+{\hat{k}_{s_1}\cdot \hat{k}_{s_2}\over 2\,\hat{s}_{s_1s_21}}\Big)\,{\cal A}_{\rm NLSM}(1,\cdots,n)\,,\nn
& &{\cal A}^{(1)}_{\rm NLSM}(1,\cdots,n,s_1,s_2)\,\xrightarrow[B=\{2,\cdots,n-2\}]{k_a\cdot k_b=0}\nn
& &~~~~~~~~~~~~~~~~~~~~~~~~\tau\,\Big[{(\hat{k}_{s_1}-\hat{k}_{s_2})\cdot k_n\over 2\,\hat{s}_{ns_1s_2}}+{(\hat{k}_{s_2}-\hat{k}_{s_1})\cdot k_1\over 2\,\hat{s}_{s_1s_21}}\Big]\,(k_{s_1}+k_{s_2})\cdot\partial_{k_n}\,{\cal A}_{\rm NLSM}(1,\cdots,n)\nn
& &~~~~~~~~~~~~~~~~~~~~~~~~+\tau\,\Big({\hat{k}_{s_1}\cdot \hat{k}_{s_2}\over 2\,\hat{s}_{ns_1s_2}}+{\hat{k}_{s_1}\cdot \hat{k}_{s_2}\over 2\,\hat{s}_{s_1s_21}}\Big)\,{\cal A}_{\rm NLSM}(1,\cdots,n)\,.~~\label{2split-NLSM-sub}
\eea
Comparing \eref{2split-NLSM-sub} with \eref{NLSM-subleading-step1}, we find
\bea
{\cal R}^{(0)}\,&\xrightarrow[B=\{3,\cdots,n-1\}]{k_a\cdot k_b=0}&\,\tau\,{(\hat{k}_{s_1}-\hat{k}_{s_2})\cdot k_n\over 2\,\hat{s}_{ns_1s_2}}\,(\hat{k}_{s_1}+\hat{k}_{s_2})\cdot(\partial_{k_1}-\partial_{k_n})\,{\cal A}_{\rm YM}(1,\cdots,n)\,,\nn
{\cal R}^{(0)}\,&\xrightarrow[B=\{2,\cdots,n-2\}]{k_a\cdot k_b=0}&\,\tau\,{(\hat{k}_{s_1}-\hat{k}_{s_2})\cdot k_1\over 2\,\hat{s}_{s_1s_21}}\,(\hat{k}_{s_1}+\hat{k}_{s_2})\cdot(\partial_{k_1}-\partial_{k_n})\,{\cal A}_{\rm YM}(1,\cdots,n)\,.~~\label{R0-constrain-NLSM}
\eea
The observation at the end of the previous subsection also makes sense for the above ${\cal R}^{(0)}$, namely, the replacement \eref{replace} converts \eref{R0-constrain-NLSM} for the NLSM case to \eref{R0-constrain-YM} for the YM case.
In practice, the above observation greatly simplifies the construction for the full unique ${\cal R}^{(0)}$. One can repeat the process from \eref{defin-effc-partial} until \eref{R0-result-YM}, to obtain
\bea
{\cal R}^{(0)}\,=\,{\tau\over 4}\,(\hat{k}_{s_1}-\hat{k}_{s_2})\cdot(\partial_{k_1}-\partial_{k_n})\,{\cal A}_{\rm YM}(1,\cdots,n)\,.~~\label{R0-result-NLSM}
\eea
Notice that the solution \eref{solu-c} is obtained by employing the observation that $\hat{k}_s\cdot k_n$ in the numerator cancels $\hat{s}_{ns}$ in the denominator. In the current NLSM case, $(\hat{k}_{s_1}+\hat{k}_{s_2})\cdot k_n$ can not cancel $\hat{s}_{ns_1s_2}$ in general. Fortunately, at the sub-leading order under consideration, $\hat{s}_{ns_1s_2}$ only contributes $2(\hat{k}_{s_1}+\hat{k}_{s_2})\cdot k_n$ to ${\cal R}^{(0)}$, thus the cancellation effectively holds. The extremely similar argument holds for $(\hat{k}_{s_1}+\hat{k}_{s_2})\cdot k_1$ and $\hat{s}_{s_1s_21}$.
Substituting ${\cal R}^{(0)}$ into \eref{NLSM-subleading-step1}, we finally obtain
\bea
{\cal A}^{(1)}_{\rm NLSM}(1,\cdots,n,s_1,s_2)\,=\,\tau\,S^{(1)}_{\rm NLSM}\,{\cal A}_{\rm NLSM}(1,\cdots,n)\,,~~\label{softtheo-NLSM-sub}
\eea
where the soft factor is given by
\bea
S^{(1)}_{\rm NLSM}&=&{k_n\cdot \hat{L}_{s_1,s_2}\cdot\partial_{k_n}\over \hat{s}_{ns_1s_2}}+{k_1\cdot \hat{L}_{s_2,s_1}\cdot\partial_{k_1}\over \hat{s}_{s_1s_21}}\,+\,\Big({\hat{k}_{s_1}\cdot \hat{k}_{s_2}\over 2\,\hat{s}_{ns_1s_2}}+{\hat{k}_{s_1}\cdot \hat{k}_{s_2}\over 2\,\hat{s}_{s_1s_21}}\Big)\,,~~\label{softfac-NLSM-sub}
\eea
where $\hat{L}^{\mu\nu}_{s_1,s_2}\equiv \hat{k}_{s_1}^\mu \hat{k}_{s_2}^\nu-\hat{k}_{s_2}^\mu \hat{k}_{s_1}^\nu$, $\hat{L}^{\mu\nu}_{s_2,s_1}\equiv \hat{k}_{s_2}^\mu \hat{k}_{s_1}^\nu-\hat{k}_{s_1}^\mu \hat{k}_{s_2}^\nu$. As shown in \cite{Zhou:2023quv,Zhou:2023vzl,Zhou:2024qjh}, the above sub-leading soft factor \eref{softfac-NLSM-sub}
is equivalent to the standard one in \cite{Cachazo:2015ksa,Du:2015esa}.

\subsection{Verification and remarks}
\label{subsec-veri-NLSM}

As in ${\rm Tr}(\phi^3)$ and YM cases, logically the sub-leading soft theorem found in \eref{softtheo-NLSM-sub} and \eref{softfac-NLSM-sub} is restricted in the subspace, defined by $k_a\cdot k_b=0$ with $a\in\{s_1,s_2\}$ and $b\in\{3,\cdots,n-2\}$. Although this sub-leading soft theorem is equivalent to the standard one in the literature, in order to make our method to be self contained, we need to give a verification for the correctness, without knowing the standard result. The argument based on $2$-splits in section \ref{subsec-veri-YM-sub} is a general one, it is natural to expect that such method can be applied to any theory satisfying $2$-split. Thus, we can apply this method to show the correctness of sub-leading soft theorem \eref{softtheo-NLSM-sub}.

Choosing $A=\{s_1,s_2\}$, $B=\{2,\cdots,n-1\}\setminus k$, the $2$-split \eref{2split-NLSM-general} is specialized to
\bea
{\cal A}_{\rm NLSM}(1,\cdots,n,s_1,s_2)\,&&\xrightarrow[B=\{2,\cdots,n-1\}\setminus k]{k_a\cdot k_b=0}\nn
&&{\cal J}_{{\rm NLSM}\oplus{\rm Tr}(\phi^3)}(n_\phi,s_{1p},s_{2p},1_\phi,\kappa_\phi)\,\times\,{\cal J}_{\rm NLSM}(1,\cdots,\kappa',\cdots,n)\,,
\eea
where the $5$-point current ${\cal J}_{{\rm NLSM}\oplus{\rm Tr}(\phi^3)}(n_\phi,s_{1p},s_{2p},1_\phi,\kappa_\phi)$ is the same as that in \eref{J-5p-NLSM+phi}. In the double-soft limit, the sub-leading behavior of this $2$-split is given by
\bea
&&{\cal A}^{(1)}_{\rm NLSM}(1,\cdots,n,s_1,s_2)\,\xrightarrow[B=\{2,\cdots,n-1\}\setminus k]{k_a\cdot k_b=0}\,{\cal J}^{(1)}_{{\rm NLSM}\oplus{\rm Tr}(\phi^3)}(n_\phi,s_{1p},s_{2p},1_\phi,\kappa_\phi)\,\times\,{\cal J}^{(0)}_{\rm NLSM}(1,\cdots,\kappa',\cdots,n)\nn
&&~~~~~~~~~~~~~~~~~~~~~~~~~~~~~~~~~~~~~~~~+{\cal J}^{(0)}_{{\rm NLSM}\oplus{\rm Tr}(\phi^3)}(n_\phi,s_{1p},s_{2p},1_\phi,\kappa_\phi)\,\times\,{\cal J}^{(1)}_{\rm NLSM}(1,\cdots,\kappa',\cdots,n)\,,~~\label{veri-NLSM-sub-1}
\eea
Using the expression of $5$-point current given in \eref{J-5p-NLSM+phi}, and ${\cal J}^{(0)}_{\rm NLSM}(1,\cdots,\kappa',\cdots,n)={\cal A}_{\rm NLSM}(1,\cdots,n)$, one can turn the above ${\cal A}^{(1)}_{\rm NLSM}(1,\cdots,n,s_1,s_2)$ to
\bea
{\cal A}^{(1)}_{\rm NLSM}(1,\cdots,n,s_1,s_2)\,&&\xrightarrow[B=\{2,\cdots,n-1\}\setminus k]{k_a\cdot k_b=0}\,\tau\,\Big({\hat{k}_{s_1}\cdot \hat{k}_{s_2}\over 2\,\hat{s}_{ns_1s_2}}+{\hat{k}_{s_1}\cdot \hat{k}_{s_2}\over 2\,\hat{s}_{s_1s_21}}\Big)\,\times\,{\cal A}_{\rm NLSM}(1,\cdots,n)\nn
&&~~~~~~~~+\Big({(\hat{k}_{s_1}-\hat{k}_{s_2})\cdot k_n\over 2\,\hat{s}_{ns_1s_2}}+{(\hat{k}_{s_2}-\hat{k}_{s_1})\cdot k_1\over 2\,\hat{s}_{s_1s_21}}\Big)\,\times\,{\cal J}^{(1)}_{\rm NLSM}(1,\cdots,\kappa',\cdots,n)\,,~~\label{veri-NLSM-sub-2}
\eea
where ${\cal J}^{(1)}_{\rm YM}(1,\cdots,\kappa',\cdots,n)$ is at the $\tau$ order.
This is the constraint which can be imposed to sub-leading double-soft behavior of NLSM amplitudes.

Then we calculate the effect of acting the operator $S^{(1)}_{\rm NLSM}$ in \eref{softfac-NLSM-sub} on ${\cal A}_{\rm NLSM}(1,\cdots,n)$.
A few algebra ultimately yields
\bea
&&S^{(1)}_{\rm NLSM}\,{\cal A}_{\rm NLSM}(1,\cdots,n)\,\xrightarrow[B=\{2,\cdots,n-1\}\setminus k]{k_a\cdot k_b=0}\,\tau\,\Big({\hat{k}_{s_1}\cdot \hat{k}_{s_2}\over 2\,\hat{s}_{ns_1s_2}}+{\hat{k}_{s_1}\cdot \hat{k}_{s_2}\over 2\,\hat{s}_{s_1s_21}}\Big)\,\times\,{\cal A}_{\rm NLSM}(1,\cdots,n)\nn
&&~~~~~~~~~~~~~~~~+\tau\,\Big({(\hat{k}_{s_1}-\hat{k}_{s_2})\cdot k_n\over 2\,\hat{s}_{ns_1s_2}}+{(\hat{k}_{s_2}-\hat{k}_{s_1})\cdot k_1\over 2\,\hat{s}_{s_1s_21}}\Big)\,\times\nn
&&~~~~~~~~~~~~~~~~~~~~\Big[(\hat{k}_{s1}+\hat{k}_{s2})\cdot k_1\,(\partial_{k_n\cdot k_1}-\partial_{k_n\cdot k_k})+(\hat{k}_{s1}+\hat{k}_{s_2})\cdot k_n\,(\partial_{k_1\cdot k_n}-\partial_{k_1\cdot k_k})\Big]\,{\cal A}_{\rm YM}(1,\cdots,n)\,.~~\label{veri-NLSM-sub-3}
\eea
The derivation of second and third lines in the above is the same as that in Appendix \ref{appen}, with the replacement \eref{replace}.
Comparing \eref{veri-NLSM-sub-3} with \eref{veri-NLSM-sub-2}, we see that the soft factor in \eref{softfac-NLSM-sub} already furnishes the required feature of sub-leading soft behavior in \eref{veri-NLSM-sub-2}. If ${\cal R}^{(0)}$ found in \eref{R0-result-NLSM} is not the complete one, the missed part can only alter the expression of ${\cal J}^{(1)}_{\rm YM}(1,\cdots,\kappa',\cdots,n)$ in \eref{veri-NLSM-sub-3}, i.e., the missed part should behave as the factorized form
\bea
\Big({(\hat{k}_{s_1}-\hat{k}_{s_2})\cdot k_n\over 2\,\hat{s}_{ns_1s_2}}+{(\hat{k}_{s_2}-\hat{k}_{s_1})\cdot k_1\over 2\,\hat{s}_{s_1s_21}}\Big)\,\times\,\cdots\,.~~\label{factorize-miss}
\eea

Instead of $\hat{k}_{s_1}$ and $\hat{k}_{s_2}$, we can choose $\hat{k}_{s_1}+\hat{k}_{s_2}$ and $\hat{k}_{s_1}-\hat{k}_{s_2}$ as two independent momenta, then the potential undetected terms in ${\cal R}^{(0)}$ are those proportional to $(\hat{k}_{s_1}+\hat{k}_{s_2})\cdot V$ or $(\hat{k}_{s_1}-\hat{k}_{s_2})\cdot V$, where $V$ are combinations of $k_b$ with $b\in\{3,\cdots,n-2\}$. Momenta $\hat{k}_{s_1}+\hat{k}_{s_2}$ and $\hat{k}_{s_1}-\hat{k}_{s_2}$ do not enter the remaining part of each undetected term, since the sub-leading order is linear in $\tau$, and the ansatz \eref{NLSM-subleading-step1} forbids ${\cal R}^{(0)}$ to have denominators depend on $\tau$. The candidates $(\hat{k}_{s_1}+\hat{k}_{s_2})\cdot V$ and $(\hat{k}_{s_1}-\hat{k}_{s_2})\cdot V$ can never be turned to
the factorized form \eref{factorize-miss}. For the first one $(\hat{k}_{s_1}+\hat{k}_{s_2})\cdot V$, since $\hat{k}_{s_1}+\hat{k}_{s_2}$ and $\hat{k}_{s_1}-\hat{k}_{s_2}$ are independent, it is impossible to create the factor in the bracket in \eref{factorize-miss}.
For the second one $(\hat{k}_{s_1}-\hat{k}_{s_2})\cdot V$, the detectable part on the locus $k_a\cdot k_b=0$ with $B=\{2,\cdots,n-1\}\setminus k$ is $(\hat{k}_{s_1}-\hat{k}_{s_2})\cdot k_k$, which is equivalent to $-(\hat{k}_{s_1}-\hat{k}_{s_2})\cdot(k_1+k_n)$. Thus, $(\hat{k}_{s_1}-\hat{k}_{s_2})\cdot k_1$ and $(\hat{k}_{s_1}-\hat{k}_{s_2})\cdot k_n$ arise from this candidate must have the same coefficient,
but this requirement can not be satisfied by the form \eref{factorize-miss}. Consequently, the missed terms in ${\cal R}^{(0)}$ can not exist.

Similar as in the YM case, the result in \eref{veri-NLSM-sub-3} also gives rise to the sub-leading double-soft behavior of the current ${\cal J}_{\rm YM}(1,\cdots,\kappa'\cdots,n)$, namely,
\bea
& &{\cal J}^{(1)}_{\rm YM}(1,\cdots,\kappa',\cdots,n)\nn
&=&\Big[(\hat{k}_{s1}+\hat{k}_{s2})\cdot k_1\,(\partial_{k_n\cdot k_1}-\partial_{k_n\cdot k_k})+(\hat{k}_{s1}+\hat{k}_{s_2})\cdot k_n\,(\partial_{k_1\cdot k_n}-\partial_{k_1\cdot k_k})\Big]\,{\cal A}_{\rm YM}(1,\cdots,n)\,.~~~\label{softtheo-NLSMcurrent}
\eea
Similar as emphasized at the end of section \ref{subsec-veri-YM-sub}, the statement of sub-leading double-soft behavior of the current corresponds to taking $k_{s_1}$ and $k_{s_2}$ in $k_{\kappa'}=k_k+k_{s_1}+k_{s_2}$ to be soft.

Further remarks are in order. The first one is about the specific behavior of NLSM amplitudes in the single-soft limit, called Adler zero. The double-soft factors $S^{(0)}_{\rm NLSM}$ and $S^{(1)}_{\rm NLSM}$ given in \eref{softtheo-NLSM-leading} and \eref{softfac-NLSM-sub} manifestly vanish when $\hat{k}_{s_2}=0$ or $\hat{k}_{s_1}=0$, this behavior is known as Adler zero, which states that the tree-level NLSM amplitudes vanish when taking one of external momenta to be soft.

We have observed that the leading soft theorems \eref{softtheo-YM-leading} and \eref{softtheo-NLSM-leading} are connected by the replacement \eref{replace}, while \eref{R0-constrain-NLSM} is related to \eref{R0-constrain-YM} via the same replacement. Such replacement also links the Adler zero of NLSM amplitudes and the gauge invariance of YM amplitudes together. Suppose we substitute $\epsilon_s=k_s$ or $\epsilon_s=-k_s$ in YM amplitudes, the replacement \eref{replace} translates them to
\bea
\hat{k}_{s_1}-\hat{k}_{s_2}=\hat{k}_{s_1}+\hat{k}_{s_2}~~~~{\rm or}~~~~\hat{k}_{s_2}-\hat{k}_{s_1}=\hat{k}_{s_1}+\hat{k}_{s_2}\,,
\eea
which forces $\hat{k}_{s_2}=0$ or $\hat{k}_{s_1}=0$ respectively. The substitution $\epsilon_s=k_s$ or $\epsilon_s=-k_s$ annihilates YM amplitudes, due to the gauge invariance. Meanwhile, setting $\hat{k}_{s_2}=0$ or $\hat{k}_{s_1}=0$ also annihilates NLSM amplitudes, thus the replacement \eref{replace} transmutes the gauge invariance to the Adler zero.

As discussed at the end of subsection \ref{subsec-leading-NLSM}, the the expression of leading soft theorem in \eref{softtheo-NLSM-leading} also contains terms at higher orders. The formula \eref{softfac-NLSM-sub} should be understood similarly, it also involves contribution at higher orders, due to the expression \eref{propagat-diver-NLSM}. To keep the compaction, we have not expanded $1/\hat{s}_{ns_1s_2}$ or $1/\hat{s}_{s_1s_21}$ until now. However,
as pointed out below \eref{R0-result-NLSM}, $(\hat{k}_{s_1}+\hat{k}_{s_2})\cdot k_n$ (or $(\hat{k}_{s_1}+\hat{k}_{s_2})\cdot k_1$) only cancels the leading contribution of $1/\hat{s}_{ns_1s_1}$ (or $1/\hat{s}_{s_1s_21}$).
It means, the expansion of $1/\hat{s}_{ns_1s_1}$ and $1/\hat{s}_{s_1s_21}$ can not be avoid when studying soft behavior at $m^{\rm th}$ order with $m>1$.

Such expansion brings new complexity into the investigation of the soft behavior at $m^{\rm th}$ order, and such complexity grows fast with increasing $m$. One can still apply the approach in this subsection to seek $S^{(m)}_{\rm NLSM}$ at higher orders, in the lower-dimensional kinematic space. However, an universal representation of $S^{(m)}_{\rm NLSM}$ which is valid for arbitrary $m$, just like \eref{softfac-phi3-high} and \eref{softfac-YM-high} in previous cases, is far from obvious.

\section{Summary and discussion}
\label{sec-summary}

In this paper, we extend the method in \cite{Arkani-Hamed:2024fyd} for constructing soft theorems through both the conventional factorization and the novel 2-split behavior. Our approach yields complete leading and sub-leading soft factors for:
(i) single-soft theorems of tree-level ${\rm Tr}(\phi^3)$ and YM amplitudes, and
(ii) double-soft theorems of NLSM amplitudes.
Furthermore, we propose a self-contained verification method based on 2-splits that requires no prior knowledge of standard soft theorems. The same technique allows us to derive universal representations of higher order ($m^{\rm th}$ order) soft theorems for ${\rm Tr}(\phi^3)$ and YM amplitudes, valid for arbitrary positive integer m. However, for $m\geq2$, these theorems are restricted to lower-dimensional kinematic subspaces rather than the full space. Crucially, all obtained soft factors maintain the consistency with momentum conservation, their action on lower-point amplitudes produces correct soft behavior under any momentum-conserving re-parameterization.

Our investigation also yields two interesting by-products:
\begin{enumerate}
\setlength{\itemindent}{-0em}
\item A simple kinematic replacement connects single-soft YM theorems with double-soft NLSM theorems at leading/sub-leading orders, this replacement also converts gauge invariance to Adler zero.
\item Universal sub-leading soft theorems for pure YM and NLSM currents in 2-splits, when the specific components of the combinatorial off-shell momentum $k_{\kappa'}$ go soft. These soft theorems may illuminate higher-order soft behaviors in general 2-splits \eref{2split-phi3-general}, \eref{2split-YM-general} and \eref{2split-NLSM-general}. The importance of such higher order soft behaviors will be pointed out in the last paragraph of this section
\end{enumerate}

In principle, for each tree-level ordered amplitude, any multi-soft theorem for arbitrary multiplicity can also be derived via our method. At least, the leading soft theorem can be fully determined by the resulting current in the corresponding 2-split that carries soft particles. However, for a large number of soft particles, the calculation of corresponding currents becomes a formidable challenge since their complexity grows rapidly with the increasing number of external legs. On the other hand, since one motivation of this paper is the method of constructing amplitudes from soft theorems, we expect that each recursive construction employs only lowest-point on-shell amplitudes as input. Each lowest-point building block corresponds to the simplest soft theorems with the least number of soft particles. For the above reasons, in this paper we consider only single-soft theorems of amplitudes whose lowest-point interactions are cubic, and double-soft theorems of amplitudes whose lowest-point interactions are quartic.

The leading and sub-leading soft theorems of YM amplitudes can also be derived by simultaneously imposing both the usual factorization and gauge invariance. This follows from the well known conclusion that locality, unitarity, and gauge invariance completely determine any tree-level YM amplitude. An insightful discussion of gluon soft theorems from this perspective can be found in the recent excellent work \cite{Backus:2025njt}, within the framework of scaffolded gluons.
Similarly, we know that usual factorizations on physical poles, when combined with Adler zeros, fully determine all tree-level NLSM amplitudes. Therefore, one would naturally expect that NLSM soft theorems could be derived by combining usual factorization with Adler zero. However, our purpose is to develop a general method applicable to any tree-level amplitude exhibiting $2$-split behavior. Hence, the constructions presented in this paper remain independent of any theory-specific properties, such as gauge invariance or Adler zero.

In the method presented in this paper, we exclusively employ the special $2$-splits in \eref{2split-phi3-special}, \eref{2split-YM-special}, and \eref{2split-NLSM-special} for our construction. The rationale for this choice lies in the fact that these particular splittings make the soft behaviors of the resulting pure ${\rm Tr}(\phi^3)$/YM/NLSM currents beyond leading order more tractable.
However, to extend our method and broaden its applicability, consideration of more general $2$-splits becomes essential. This necessity arises from two key observations:
\begin{enumerate}
\setlength{\itemindent}{-0.2em}
\item As demonstrated, while the leading soft theorem can be determined using just one $2$-split, establishing the full sub-leading theorem requires two distinct $2$-splits. Thus, to obtain complete higher order soft behaviors rather than partial results in subspaces, employing additional $2$-splits appears both natural and necessary.

\item Various unordered amplitudes, including gravitational, special Galileon, and Born-Infeld cases, exhibit remarkably similar $2$-split behaviors \cite{Cao:2024qpp}. In these unordered scenarios, the concept of "nearby legs in the ordering" becomes meaningless, and thus no direct analogues exist for the special $2$-splits \eref{2split-phi3-special}, \eref{2split-YM-special}, and \eref{2split-NLSM-special}. Therefore, generalizing our approach to such unordered amplitudes necessarily requires analyzing soft behaviors of currents in general $2$-splits.
\end{enumerate}

\appendix

\section{Derivation of \eref{veri-YM-sub-3}}
\label{appen}

In this section we give the detailed derivation of the factorization formula in \eref{veri-YM-sub-3}.

According to momentum conservation, we have
\bea
\{\hat{k}_s,\epsilon_s\}\,\cdot\,(k_1+k_k+k_n)\,\xrightarrow[B=\{2,\cdots,n-1\}\setminus k]{k_s\cdot k_b=0\,,~\eref{condi-polar}}&\,\{\hat{k}_s,\epsilon_s\}\,\cdot\,\Big(\sum_{a=1}^n\,k_a\Big)=0\,,
\eea
thus \eref{ke-dot-remain} can be converted to
\bea
&&\{\hat{k}_s,\epsilon_s\}\,\cdot\,\partial^\mu_{k_n}\,{\cal A}_{\rm YM}(1,\cdots,n)\,\xrightarrow[B=\{2,\cdots,n-1\}\setminus k]{k_s\cdot k_b=0\,,~\eref{condi-polar}}\nn
&&~~~~~~~~~~~~~~~~~~~~~~~~~~~~~~~~~~~~\{\hat{k}_s,\epsilon_s\}\,\cdot\,\Big[k_n\,(\partial_{k_n\cdot k_n}-\partial_{k_n\cdot k_k})+k_1\,(\partial_{k_n\cdot k_1}-\partial_{k_n\cdot k_k})\Big]\,{\cal A}_{\rm YM}(1,\cdots,n)\,,\nn
&&\{\hat{k}_s,\epsilon_s\}\,\cdot\,\partial^\mu_{k_1}\,{\cal A}_{\rm YM}(1,\cdots,n)\,\xrightarrow[B=\{2,\cdots,n-1\}\setminus k]{k_s\cdot k_b=0\,,~\eref{condi-polar}}\nn
&&~~~~~~~~~~~~~~~~~~~~~~~~~~~~~~~~~~~~\{\hat{k}_s,\epsilon_s\}\,\cdot\,\Big[k_1\,(\partial_{k_1\cdot k_1}-\partial_{k_1\cdot k_k})+k_n\,(\partial_{k_1\cdot k_n}-\partial_{k_1\cdot k_k})\Big]\,{\cal A}_{\rm YM}(1,\cdots,n)\,.~~\label{ke-dot-remain-2}
\eea
Using \eref{ke-dot-remain-2}, and the expression of $S^{(1)}_{\rm YM}$ in \eref{softfac-YM-sub}, we obtain
\bea
S^{(1)}_{\rm YM}\,{\cal A}_{\rm YM}(1,\cdots,n)&&\,\xrightarrow[B=\{2,\cdots,n-1\}\setminus k]{k_s\cdot k_b=0\,,~\eref{condi-polar}}\,
\Big[{\epsilon_s\cdot k_n\over\hat{s}_{ns}}\,\hat{k}_s\cdot k_1\,(\partial_{k_n\cdot k_1}-\partial_{k_n\cdot k_k})-{\epsilon_s\cdot k_1\over\hat{s}_{s1}}\,\hat{k}_s\cdot k_n\,(\partial_{k_1\cdot k_n}-\partial_{k_1\cdot k_k})\nn
&&~~~~~~~~~~+{1\over2}\,\epsilon_s\cdot k_1\,(\partial_{k_n\cdot k_k}-\partial_{k_n\cdot k_1})+{1\over2}\,\epsilon_s\cdot k_n\,(\partial_{k_1\cdot k_n}-\partial_{k_1\cdot k_k})\Big]\,{\cal A}_{\rm YM}(1,\cdots,n)\,.~~\label{appen-1}
\eea
Inserting
\bea
{1\over2}={\hat{k}_s\cdot k_n\over\hat{s}_{ns}}={\hat{k}_s\cdot k_1\over\hat{s}_{s1}}\,,
\eea
one can turn \eref{appen-1} to
\bea
S^{(1)}_{\rm YM}\,{\cal A}_{\rm YM}(1,\cdots,n)&&\,\xrightarrow[B=\{2,\cdots,n-1\}\setminus k]{k_s\cdot k_b=0\,,~\eref{condi-polar}}\,
\Big[{\epsilon_s\cdot k_n\over\hat{s}_{ns}}\,\Big(\hat{k}_s\cdot k_1\,(\partial_{k_n\cdot k_1}-\partial_{k_n\cdot k_k})+\hat{k}_s\cdot k_n\,(\partial_{k_1\cdot k_n}-\partial_{k_1\cdot k_k})\Big)\nn
&&-{\epsilon_s\cdot k_1\over\hat{s}_{s1}}\,\Big(\hat{k}_s\cdot k_n\,(\partial_{k_1\cdot k_n}-\partial_{k_1\cdot k_k})
+\hat{k}_s\cdot k_1(\partial_{k_n\cdot k_1}-\partial_{k_n\cdot k_k})\Big)\Big]\,{\cal A}_{\rm YM}(1,\cdots,n)\,,
\eea
which is equivalent to \eref{veri-YM-sub-3}.

\bibliographystyle{JHEP}

\bibliography{reference}

\end{document}